\documentclass[fleqn,usenatbib]{mnras} 
\usepackage{hyperref} 
\usepackage{ae,aecompl}
\usepackage{ulem}
\usepackage{xargs} 
\usepackage{siunitx}
\usepackage{longtable}
\usepackage{booktabs}
\usepackage{pdflscape}

\usepackage{amsmath}	
\usepackage{natbib}
\usepackage{color}
\usepackage{lscape}
\usepackage{graphicx} 
\usepackage{amssymb}

\newcommand{\prot}{$P_{\rm rot}$}

\newcommand{\teff}{$T_{\rm eff}$}
\newcommand{\logg}{$\log g$}
\newcommand{\vsini}{$v\sin i$}

\newcommand{\ROTFIT}{{\sf ROTFIT}}

\newcommand{\kms}{km\,s$^{-1}$}
 
\newcommand{\gaia}{{\it Gaia}}

\newcommand{\Msun}{{$M_{\odot}$}}

\newcommand{\Rsun}{{$R_{\odot}$}}

\title[Rotation and activity in the ASCC\,123 open cluster]{Rotation and activity in late-type members of the young cluster ASCC 
\,123\thanks{Based on observations made with the Italian Telescopio Nazionale Galileo (TNG) operated on the island of La Palma by the Fundaci\'on Galileo Galilei of the INAF (Istituto Nazionale di Astrofisica) at the Observatorio del Roque de los Muchachos, as part of the Large Program ``Stellar Population Astrophysics" (SPA). Based on photometry collected at the INAF - Osservatorio Astrofisico di Catania.}} 
\author[A. Frasca et al.]{A. Frasca$^{1}$\thanks{E-mail: antonio.frasca@inaf.it}, 
J. Alonso-Santiago$^{1}$,
G. Catanzaro$^{1}$, 
A. Bragaglia$^{2}$
~\\  
$^{1}$INAF--Osservatorio Astrofisico di Catania, Via S.Sofia 78, I-95123, Catania, Italy\\ 
$^{1}$INAF--Osservatorio di Astrofisica e Scienza dello Spazio, Via P. Gobetti 93/3, 40129 Bologna, Italy\\
} 
 
\date{April 2023} 
 
\pubyear{2022}

\begin{document} 
\label{firstpage} 
\pagerange{\pageref{firstpage}--\pageref{lastpage}}

\maketitle 
 
\begin{abstract} 
{ASCC~123 is a little-studied young and dispersed open cluster. Recently, we conducted the first research devoted to it. In this paper, we complement our previous 
work with {\it TESS} photometry for the 55 likely members of the cluster. We pay special attention to seven of these high-probability members, all with FGK spectral 
types, for which we have high-resolution spectra from our preceding work. By studying the {\it TESS} light curves of the cluster members we determine the rotational 
period and the amplitude of the rotational modulation for 29 objects. The analysis of the distribution of the periods allows us to estimate a gyrochronogical age 
for ASCC~123 similar to that of the Pleiades, confirming the value obtained in our previous investigation. A young cluster age is also suggested by the distribution 
of variation amplitudes. In addition, for those stars with spectroscopic data we calculate the inclination of their rotation axis. These values appear to follow a 
random distribution, as already observed in young clusters, with no indication of spin alignment. However, our sample is too small to confirm this on more solid 
statistical grounds. Finally, for these seven stars we study the level of magnetic activity from the H$\alpha$ and \ion{Ca}{ii} H\&K lines. 
Despite the small number of data points, we find a correlation of the H$\alpha$ and \ion{Ca}{ii} flux with Rossby number. The position of these stars in flux--flux 
diagrams follows the general trends observed in other active late-type stars.
}

\end{abstract} 
 

\section{Introduction}

Stellar rotation and magnetic activity are properties strongly related to the evolution of stars, from the pre-main to the post-main sequence.
During the main sequence (MS) phase, the position of a star on the Hertzsprung-Russell (HR) diagram changes very little, so it is not an efficient indicator of stellar age. Conversely, the rotation rate and the level of magnetic activity, which is closely related to the rotation speed, change significantly over the life of a star in the main sequence due to the magnetic braking. The latter is caused by the angular momentum carried off by magnetized stellar wind.

Since the pioneering work of \citet{Skumanich1972}, who discovered the empirical law that bears his name, of the decline of the equatorial rotation speed with the inverse square root of the star's age, the rotation rate and magnetic activity have been used to estimate the age of stars in the solar neighbourhood  \citep[e.g.,][and references therein]{Barry1987,Soderblom1991,Barnes2003,Barnes2016,Stanford-Moore2020}.
Another  widely used age indicator is the atmospheric abundance of lithium \citep[e.g.,][and references therein]{Jeffries2014, Gutierrez2020}.

On the other hand, rotation and activity can also be used as additional criteria, along with the kinematics, lithium abundance, and the position in the HR diagram, for evaluating the probability for a star to be member of an open cluster. This can be of great help in the case of stars in the halo of young clusters or very dispersed clusters.
This is the case of ASCC\,123, a young and poorly studied cluster. 
ASCC~123 was discovered by \citet{Kharchenko2005} based on $Hipparcos$ proper motions and $BV$ archival photometry.
They found 24 likely members spread over a large region of more than 2 degrees on the sky. 
They found a low reddening, $E(B-V)=0.10$, placed the cluster at a distance of 250 pc, and estimated an age of about 260 Myr.
The parameters of this cluster were subsequently reanalyzed by \citet{Yen2018} on the basis of the $Gaia$ DR1/TGAS and HSOY data \citep{HSOY}.
They confirmed the low reddening, $E(B-V)=0.097$, a distance $d$=243.5\,pc, and revised the cluster age as $\tau=130$\,Myr.
\citet{Cantat2018}, from the $Gaia$ DR2 astrometric and photometric data, reported 55 cluster members and estimated a distance of $233.1\pm5.5$\,pc.

In our previous work \citep[][hereafter Paper~I]{Frasca2019}, we studied some of the brightest candidate members of ASCC\,123 using high-resolution spectra 
obtained with HARPS-N at the {\it Telescopio Nazionale Galileo} (TNG) with the aim of determining their physical parameters, radial and rotational velocities, 
and the abundances of some elements.
The chemical composition of this cluster turned out to be compatible with the Galactic trends in the solar neighbourhood. The HR and colour-magnitude diagrams 
(CMDs), as well as the lithium abundance and the H$\alpha$ emission allowed us to infer an age in the range 100--200 Myr, i.e. similar to that of the Pleiades.

In the present paper we use ground-based and {\it TESS} photometry for the members of ASCC\,123 to investigate their photospheric activity and study
the distribution of their rotation periods. We use also the HARPS-N spectra already gathered to derive the level of magnetic activity from the H$\alpha$ and \ion{Ca}{II} H\&K lines and the projected rotational velocity (\vsini)\ to derive the inclination of the rotation axes.
The paper is organised as follows. In Sect.~\ref{obs} we present the observations and the criteria followed to select our targets. 
In Sect.~\ref{Sec:results} we show the results of our work, describing the analysis carried out on both the photometric and spectroscopic data. 
We discuss our results and compare our findings with other clusters  in Sect.~\ref{Sec:discussion}. 
Finally, Sect.~\ref{Sec:conclusions} summarizes the main results and presents our conclusions.

\section{Observations and data reduction}
\label{obs}

\subsection{Spectroscopy}
\label{Subsec:Obs_spec}

In this paper we investigate the chromospheric activity for the single, late-type stars studied in Paper~I that are \textit{bona-fide} members (i. e. with a membership 
probability $Prob$=1) according to \citet{Cantat2018}. For these seven FGK stars we follow the same numbering used in our previous work (ID).
We use high-resolution spectra taken in 2018 and 2019 with GIARPS (GIANO-B \& HARPS-N; \citealt{Claudi2017}) at the 3.6m TNG telescope. The reader is referred to 
Paper~I for a description of the data and their reduction. The analysis of these spectra  with the code \ROTFIT, aimed at the determination of the atmospheric 
parameters and \vsini, can be also found in Paper~I. 

\setlength{\tabcolsep}{5pt}

\begin{table*}
\caption{Stellar parameters of the late-type members of ASCC\,123 from \citet{Frasca2019} and from the present work.} 
\begin{center}
\begin{tabular}{lrcccrcccccrcl}   
\hline
\noalign{\smallskip}
ID  & TIC &   RA    &  DEC   & \teff & err & SpT & \vsini & err & $M_*$ & $R_*$ & $P_{\rm rot}$ & err & ~~$i$ \\ 
                           &  & (J2000) & (J2000) & \multicolumn{2}{c}{(K)} & & \multicolumn{2}{c}{(\kms)} & 
			   (\Msun)  & (\Rsun) & \multicolumn{2}{c}{(days)} & ($\degr$) \\       
\noalign{\smallskip}
\hline
\noalign{\smallskip}
39   &  64837857 & 22 35 13.26  & +54 46 24.8 &  6667 &  115 &  F4V    &   49.1 &  1.9 & 1.38 & 1.36 & 1.67: & \dots  & $\sim 90$ \\   
214  & 249784843 & 22 38 34.03  & +53 35 08.7 &  5804 &   87 &  G1.5V  &  100.9 &  3.0 & 1.09 & 1.16 & 0.562 & 0.002  & 75$^{+11}_{6}$  \\
435  & 388696341 & 22 42 00.19  & +55 00 58.5 &  5758 &   79 &  G2.5V  &   11.6 &  0.7 & 1.07 & 0.96 & 3.92  & 0.02   & 69$^{+15}_{10}$ \\ 
517  & 428274538 & 22 43 26.53  & +54 11 58.4 &  5784 &   81 &  G2V    &   83.6 &  1.9 & 1.08 & 1.13 & 0.579 & 0.002  & 58$^{+2}_{3}$\\ 
554  & 361944360 & 22 44 00.20  & +54 08 38.1 &  6871 &  152 &  F4V    &   81.8 &  3.3 & 1.46 & 1.40 & 0.857 & 0.007  & 81$^{+9}_{10}$\\ 
F1   &  66539637 & 22 45 28.25  & +53 47 06.1 &  5263 &   92 &  K0V    &    6.6 &  0.6 & 0.94 & 0.92 & 5.83  & 0.05   & 56$^{+9}_{8}$\\ 
F2   &  64077901 & 22 31 17.98  & +55 02 40.7 &  5237 &   77 &  K1V    &    7.5 &  0.6 & 0.93 & 0.88 & 5.09  & 0.05   & 59$^{+10}_{8}$\\ 
\noalign{\smallskip}
\hline
\noalign{\smallskip}
\end{tabular}
\label{Tab:param}
\end{center}
\end{table*}

Table~\ref{Tab:param} reports the ID of the observed targets along with the {\it TESS} Input Catalog (TIC, \citealt{Stassun2019,Paegert2022}) identifier, 
equatorial coordinates (RA, DEC),  effective temperature (\teff),  spectral type (SpT) and \vsini, which were already derived in Paper~I. 
The masses ($M_*$) and radii ($R_*$), reported in columns 10 and 11 of Table~\ref{Tab:param}, have been derived from the comparison of the position 
of these stars in the HR diagram (Figure 8 in Paper~I) with the PARSEC isochrones ($age$=155\,Myr) and evolutionary tracks \citep{Bressan2012}. 
The rotation period (\prot) measured from the analysis of {\it TESS} light curves (Sect.~\ref{Sec:TESS}) and the inclination ($i$) of the rotation axis, 
calculated as described in Sect.~\ref{Subsec:stellar_spins}, are also quoted in the last columns of Table~\ref{Tab:param}.

\subsection{Photometry}
\label{Subsec:Obs_photo}

Space-born accurate photometry was obtained with  NASA's Transiting Exoplanet Survey Satellite ({\it TESS}; \citealt{Ricker2015}) for all the cluster members 
identified by \citet{Cantat2018}, 55 stars in total. Our targets were observed by {\it TESS} in sector 16, between 2019-09-12 and 2019-10-06, and sector 17, 
between 2019-10-08 and 2019-11-02.
The observations in two consecutive sectors allowed us to obtain nearly uninterrupted sequences (with a gap of about 1.4 days) of high-precision 
photometry lasting about 50 days, with a cadence of 30 minutes.
Being ASCC\,123 a sparse and nearby cluster, there is no severe star crowding, also taken the large pixel size of {\it TESS} (21\arcsec) into account.
Indeed, searching in the \gaia~DR3 catalogue \citep{GaiaDR3}, no star with a comparable magnitude ($\Delta G<2$\,mag) can be found within a radius 
of 21\arcsec around the position of each target. Therefore, we expect no relevant flux contamination from nearby sources.
We downloaded the {\it TESS} light curves \citep{Huang2020} from the MAST\footnote{\url{https://mast.stsci.edu/portal/Mashup/Clients/Mast/Portal.html}} archive 
and used the Simple Aperture Photometry flux (SAP) or the Pre-search 
Data Conditioning SAP flux (PDCSAP), where long term trends have been removed, whenever available. For the targets fainter than about $V=14$\,mag
we used the {\it TESS} light curves extracted with a PSF-based approach (PATHOS, \citealt{Nardiello2019}).

We have also performed multiband $BVR_{\rm C}I_{\rm C}$ photometric observations with the facility imaging 
camera\footnote{\url{https://openaccess.inaf.it/handle/20.500.12386/764}} at the 0.91\,m telescope of the {\it M. G. Fracastoro} station (Serra La Nave, Mt. Etna,
1735 m a.s.l.) of the {\it Osservatorio Astrofisico di Catania} (OACT, Italy). We observed the three G-type members in our sample, namely S\,214, S\,435, and S\,517 
during 9 nights from 22 October 2019 to 16 January 2020, collecting from 23 to 29 data points per each filter and per each star. Time series data covering about 
3--4 hours could be acquired only on 23 October, 4 and 20 November 2019. We decided to start observing  S\,214 and S\,517 because of their large \vsini, which would 
imply a rotation period short enough and a modulation amplitude large enough to be able to detect brightness variations with observations from the ground. S\,435 was 
included as a test target because of its spectral type similar to the other two fast-rotating stars. We did not consider the fast-rotating F4V stars S\,39 and S\,554 
for which we expected small variation amplitudes that are hardly detectable from the ground. We used exposure times of 60, 30, 15, and 15 sec for the $B$, $V$, 
$R_{\rm C}$, and $I_{\rm C}$ band, respectively. The data were reduced by subtracting master darks taken with the same exposure times as the science images and by 
dividing them by master flats. 
Aperture photometry with a radius of 6 pixels ($\simeq 4\arcsec$) was performed for the unsaturated stars in the field of view of 11.2$'\times 11.2'$ of each of the 
observed targets.
Standard stars in the open cluster NGC\,7790 \citep{Stetson2000} were also observed in the nights with the best photometric conditions to calculate the zero points 
and transformation coefficients to the Johnson-Cousins system. With the latter we calculated the $BVR_{\rm C}I_{\rm C}$ magnitudes for a number of non-variable stars 
in each target field that have been used as comparison for doing ensemble photometry of the monitored stars.
Although the precision and cadence from this ground-based photometry is not comparable to that of {\it TESS}, these multiband data are useful to determine average 
magnitudes and colours for these three targets. Since in the meantime {\it TESS} had started to observe the sky region containing ASCC\,123 providing 
a very precise photometry (although with no colour information), we decided not to continue with the observations from the ground. 

The main results presented in this paper are based on the {\it TESS} photometry only. 
For the sake of completeness, we report the results of ground-based photometry in Appendix~\ref{Sec:Ground_Phot}.

\section{Data analysis and results}
\label{Sec:results}

\subsection{{\it TESS} light curves}
\label{Sec:TESS}

\setlength{\tabcolsep}{3pt}

\begin{table*}
\caption{Rotation periods and variation amplitudes for the members of ASCC\,123  derived in the present work.} 
\begin{center}
\begin{tabular}{rccccrccccll}
\hline
\noalign{\smallskip}
  TIC       &  RA         &  DEC   &  Source$^a$ & $G^a$   & $G_{\rm BP}-G_{\rm RP}^a$ & $V^b$ & $K_{\rm S}^c$ & Prob$^d$ & Ampl & $P_{\rm rot}$ & err  \\
            & (J2000)     &(J2000) &             & (mag)   &      (mag)                & (mag) &  (mag)        &          & (mag)& \multicolumn{2}{c}{(days)} \\
\noalign{\smallskip}
\hline
\noalign{\smallskip}
  467546937 & 22 29 27.77 & +53 16 35.4 & 2001745344554471296 &  16.062897  &  2.862838  &  17.233  & 12.225  &  1.0 &  0.0726 &  5.40  &  0.26   \\  
  249784843 & 22 38 34.03 & +53 35 08.7 & 2003023629898711680 &  11.816078  &  0.914368  &  12.187  & 10.194  &  1.0 &  0.0836 &  0.562 &  0.002  \\  
  428062248 & 22 36 24.68 & +53 15 06.1 & 2003011500910639744 &  15.950929  &  2.474563  &  16.461  & 12.419  &  1.0 &   \dots &  \dots &  \dots  \\  
  452862919 & 22 34 35.38 & +53 05 22.0 & 2002983291564863872 &  17.446210  &  2.999275  &  17.970  & 13.511  &  1.0 &   \dots &  \dots &  \dots  \\  
   66541342 & 22 46 07.28 & +53 30 19.7 & 2002130929538238336 &  15.743083  &  2.433725  &  16.906  & 12.153  &  1.0 &  0.0285 &  6.22  &  0.81   \\  
   66541343 & 22 46 06.80 & +53 30 19.2 & 2002130929538237952 &  17.163408  &  3.006250  &  \dots   & 13.104  &  0.8 &   \dots &  \dots &  \dots  \\  
  298019363 & 22 51 06.72 & +54 28 54.2 & 2002757891681122944 &  17.132408  &  3.015108  &  17.150  & 13.098  &  0.8 &   \dots &  \dots &  \dots  \\  
  427062959 & 22 39 36.92 & +53 07 03.0 & 2002201225257443072 &  13.165830  &  1.231491  &  13.503  & 11.045  &  1.0 &  0.0461 &  5.29  &  0.02   \\  
   64073268 & 22 30 24.04 & +54 08 38.9 & 2001844163164756992 &  17.318691  &  3.011692  &  17.970  & 13.382  &  0.6 &   \dots &  \dots &  \dots  \\  
  197755998 & 22 28 01.49 & +53 47 39.9 & 2001820798541996928 &  17.113846  &  2.967771  &  \dots   & 13.178  &  0.9 &   \dots &  \dots &  \dots  \\  
  361944444 & 22 44 20.64 & +54 10 08.3 & 2002409582707022592 &  17.910720  &  3.171944  &  \dots   & 13.869  &  0.8 &   \dots &  \dots &  \dots  \\  
  361944360 & 22 44 00.20 & +54 08 38.1 & 2002409483936262016 &  10.176037  &  0.637021  &  10.410  &  9.168  &  1.0 &  0.0024 &  0.857 &  0.007  \\  
   66194421 & 22 43 13.83 & +53 53 46.0 & 2002397698545886592 &  16.360283  &  2.723582  &  17.692  & 12.595  &  0.8 &  0.2267 &  2.15  &  0.06   \\  
  428274538 & 22 43 26.53 & +54 11 58.4 & 2003161751738142464 &  11.994698  &  0.997400  &  12.199  & 10.205  &  1.0 &  0.0517 &  0.579 &  0.002  \\  
   64838038 & 22 34 51.43 & +54 43 20.3 & 2003377530902056064 &  17.930496  &  3.233290  &  \dots   & 13.856  &  0.9 &   \dots &  \dots &  \dots  \\  
   64837857 & 22 35 13.26 & +54 46 24.8 & 2003378188041736320 &  10.356592  &  0.619606  &  10.385  &  9.291  &  1.0 &  0.0051 &  1.67  &  0.01   \\  
  343437865 & 22 46 13.51 & +53 59 10.6 & 2002718377980766336 &  17.758833  &  3.240006  &  \dots   & 13.636  &  0.8 &   \dots &  \dots &  \dots  \\  
  317272583 & 22 48 38.34 & +54 44 01.6 & 2002822934667425408 &  17.820822  &  3.080757  &  \dots   & 13.742  &  0.7 &   \dots &  \dots &  \dots  \\  
  317273771 & 22 48 47.90 & +54 24 53.6 & 2002801150593176064 &   6.112235  & -0.092576  &   6.134  &  6.310  &  1.0 &   \dots &  \dots &  \dots  \\  
  367686364 & 22 55 46.32 & +54 57 08.8 & 2002876535856254464 &  10.035475  &  0.606720  &  10.255  &  9.050  &  0.8 &  0.0023 &  4.37  &  0.08   \\  
   67055292 & 22 50 14.85 & +53 33 28.1 & 2002471537620861056 &  16.707653  &  2.859225  &  17.560  & 12.819  &  0.9 &   \dots &  \dots &  \dots  \\  
   64647628 & 22 34 50.16 & +54 10 04.2 & 2003297923690672512 &  15.213570  &  2.429640  &  16.294  & 11.692  &  1.0 &  0.0588 &  2.71  &  0.08   \\  
   64278003 & 22 31 36.44 & +54 04 19.3 & 2003330050047078144 &  17.137224  &  3.127945  &  \dots   & 13.098  &  1.0 &   \dots &  \dots &  \dots  \\  
   64077901 & 22 31 17.98 & +55 02 40.7 & 2006435105245732480 &  12.769458  &  1.083085  &  12.805  & 10.839  &  1.0 &  0.0961 &  5.09  &  0.05   \\  
  415468735 & 22 29 36.22 & +55 16 16.8 & 2006455961606619520 &  13.152347  &  1.193418  &  13.399  & 11.100  &  1.0 &  0.0291 &  3.99  &  0.03   \\  
   64561694 & 22 33 59.64 & +55 42 22.4 & 2006489462352825344 &  15.635015  &  2.427858  &  16.505  & 12.067  &  0.6 &  0.1949 &  6.61  &  0.45   \\  
  317271523 & 22 48 25.68 & +55 01 22.3 & 2003593864116296576 &  16.727818  &  2.879941  &  17.800  & 12.902  &  1.0 &   \dots &  \dots &  \dots  \\  
  416319381 & 22 37 50.00 & +56 11 54.5 & 2006619303494467328 &  17.878110  &  3.284651  &  \dots   & 13.790  &  0.6 &   \dots &  \dots &  \dots  \\  
  249725057 & 22 37 01.79 & +53 34 52.1 & 2003031841875970176 &  16.084180  &  2.522219  &  17.232  & 12.480  &  1.0 &   \dots &  \dots &  \dots  \\  
  388387063 & 22 40 14.32 & +54 32 27.7 & 2003216972147204608 &  13.000790  &  1.158239  &  13.382  & 10.930  &  1.0 &  0.0534 &  3.19  &  0.03   \\  
   64646728 & 22 33 35.99 & +53 58 10.4 & 2003271397972097024 &  13.680970  &  1.375657  &  13.911  & 11.279  &  1.0 &  0.0799 &  4.03  &  0.07   \\  
  388651771 & 22 40 48.14 & +54 35 46.6 & 2003240920885298176 &  17.222970  &  2.861937  &  17.820  & 13.303  &  0.9 &   \dots &  \dots &  \dots  \\  
  420123691 & 22 32 07.17 & +53 42 26.4 & 2003259372063199744 &  13.262769  &  1.266625  &  13.594  & 11.053  &  1.0 &  0.4418 & 13.43  &  0.20   \\  
  388693014 & 22 41 34.98 & +54 35 30.0 & 2003228517019770240 &  17.371294  &  3.338816  &  \dots   & 13.166  &  0.9 &   \dots &  \dots &  \dots  \\  
  431151968 & 22 41 18.32 & +53 17 51.3 & 2002297368107185920 &  16.220058  &  2.902960  &  17.080  & 12.348  &  0.8 &  0.0787 &  0.910 &  0.015  \\  
   66343379 & 22 44 17.19 & +53 35 24.7 & 2002328875988217216 &  16.020758  &  2.592554  &  17.219  & 12.356  &  1.0 &  0.0456 &  4.98  &  0.30   \\  
   66539637 & 22 45 28.25 & +53 47 06.1 & 2002337603362057088 &  12.563967  &  1.053528  &  12.669  & 10.649  &  1.0 &  0.0483 &  5.83  &   0.05  \\  
 2046168414 & 22 45 28.58 & +53 47 07.9 & 2002337603353142016 &  17.408490  &  1.998825  &  18.077  & 10.649  &  0.6 &   \dots &  \dots &  \dots  \\  
  420304656 & 22 40 23.86 & +53 49 07.6 & 2003120764877992448 &  17.124058  &  2.955116  &  17.970  & 13.259  &  0.8 &   \dots &  \dots &  \dots  \\  
  431150096 & 22 41 28.14 & +53 52 51.6 & 2003127705545276032 &  12.836430  &  1.113888  &  12.977  & 10.868  &  1.0 &  0.0697 &  5.24  &   0.06  \\  
 2046389820 & 22 41 06.01 & +53 59 05.8 & 2003132344098378496 &  14.223695  &  \dots	 &  \dots   &  \dots  &  0.7 &   \dots &  \dots &  \dots  \\  
  452865760 & 22 34 58.37 & +53 56 32.2 & 2003104718883916288 &  17.162186  &  3.024750  &  17.610  & 13.226  &  0.7 &   \dots &  \dots &  \dots  \\  
  427060252 & 22 39 36.76 & +53 52 53.3 & 2003136501637932672 &  15.955221  &  2.871970  &  16.900  & 12.065  &  0.8 &  0.0451 &  0.367 &  0.003  \\  
   66723344 & 22 47 59.13 & +53 40 15.4 & 2002508302542111488 &  16.220280  &  2.973432  &  17.560  & 12.251  &  0.9 &  0.0292 &  0.311 &  0.002  \\  
   67224655 & 22 52 17.23 & +53 53 39.2 & 2002528402990066944 &  15.586474  &  2.455151  &  16.671  & 11.949  &  0.9 &  0.1094 &  1.737 &  0.031  \\  
  297927402 & 22 50 32.54 & +54 02 14.8 & 2002543967951224960 &  16.304932  &  2.683093  &  17.568  & 12.476  &  0.9 &  0.0515 &  1.503 &  0.023  \\  
  343771372 & 22 46 34.54 & +54 46 05.2 & 2003554006818945280 &   9.578720  &  0.443725  &   9.703  &  8.837  &  0.9 &         &  \dots &  \dots  \\  
 2015953058 & 22 31 48.85 & +54 59 22.7 & 2006387242128323840 &  14.283266  &  1.744160  &  14.416  & 10.775  &  1.0 &  0.0965 &  5.19  &   0.04  \\  
   64077487 & 22 30 51.98 & +54 55 50.1 & 2006385764659269760 &  16.194597  &  2.755547  &  17.110  & 12.358  &  0.6 &  0.0445 &  3.52  &   0.26  \\  
   67062121 & 22 50 59.10 & +53 19 00.0 & 2002452708484342656 &  17.461884  &  2.946167  &  17.970  & 13.457  &  0.8 &   \dots &  \dots &  \dots  \\  
  388696341 & 22 42 00.19 & +55 00 58.5 & 2003443437181978240 &  12.048476  &  0.941633  &  12.261  & 10.418  &  1.0 &  0.0838 &  3.92  &   0.02  \\  
  343867253 & 22 47 52.68 & +55 33 18.5 & 2003669180664786944 &  15.195872  &  2.754218  &  15.750  & 11.370  &  0.8 &  0.0305 &  0.464 &  0.003  \\  
  343778985 & 22 47 36.61 & +55 10 30.5 & 2003645060128342656 &  16.949617  &  2.980359  &  17.800  & 12.994  &  0.6 &   \dots &  \dots &  \dots  \\  
  431153083 & 22 41 11.77 & +52 56 27.8 & 2002274106564094848 &  13.922489  &  1.505455  &  14.256  & 11.335  &  1.0 &  0.0780 &  6.90  &   0.03  \\  
  249801592 & 22 39 19.22 & +53 29 16.5 & 2002270842392282496 &  10.599697  &  0.758198  &  10.765  &  9.314  &  1.0 &   \dots &  \dots &  \dots  \\  
\noalign{\smallskip}
\hline
\end{tabular}
\end{center}
{\bf Notes.} $^a$ From \gaia~DR3 catalogue \citep{GaiaDR3}. $^b$ From {\it TESS} Input Catalog \citep{Paegert2022}.
$^c$ From 2MASS catalogue \citep{Cutri2006}. $^d$ Membership probability according to \citet{Cantat2018}.

\label{Tab:periods}
\end{table*}

\begin{figure}
\hspace{-0.5cm}
\includegraphics[width=9.3cm]{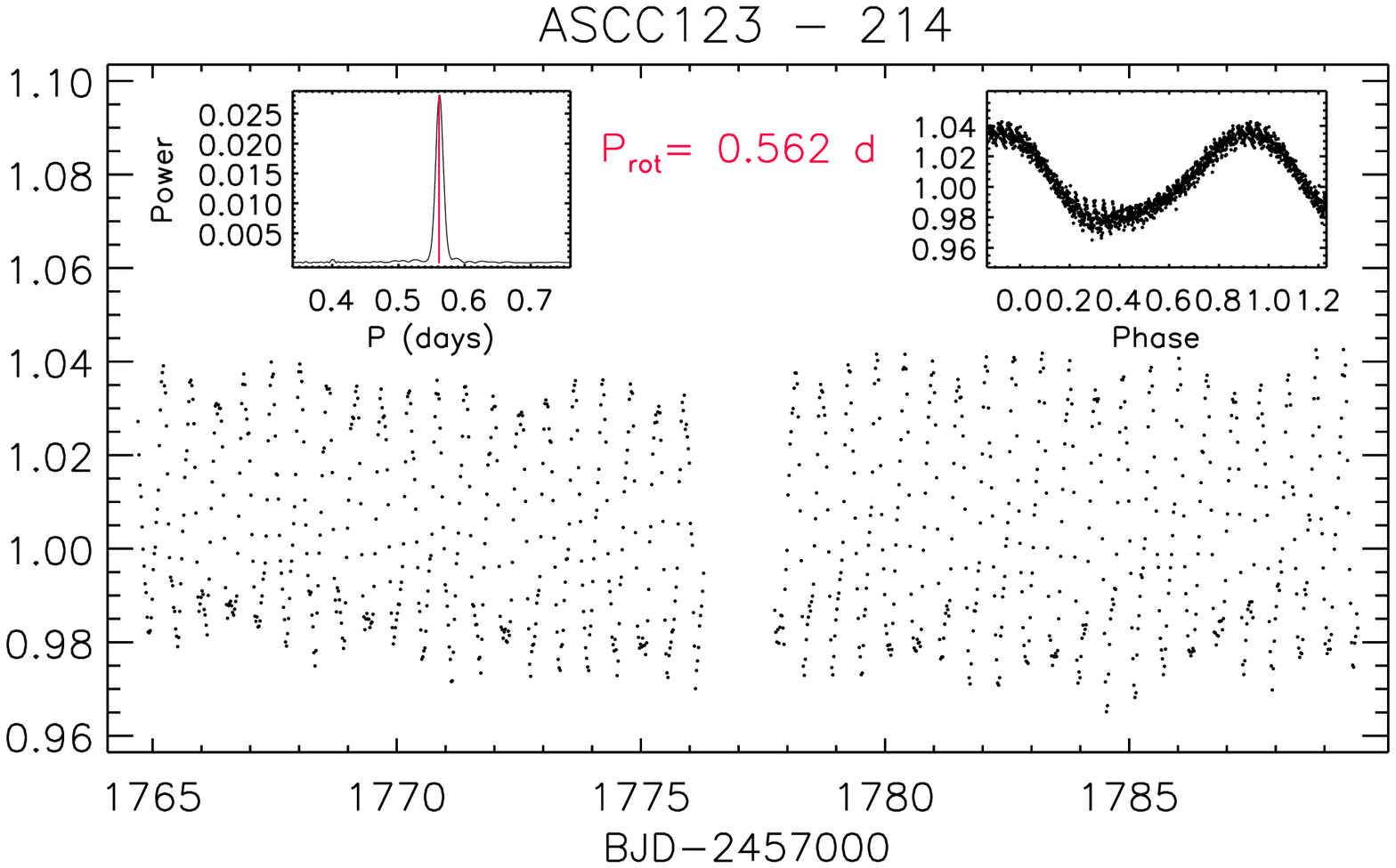}	
\vspace{-0.4cm}
\caption{{\it TESS} light curve of S\,214 (TIC\,249784843) in 2019 (sector 17). The inset in the upper left corner shows the 
cleaned periodogram of these data; the rotational period is marked with a vertical red line. The inset in the upper right corner
displays the data phased with this period.}
\label{fig:TESS_ASC214}
\end{figure}

The main information we can get from the {\it TESS} light curves is the rotation period, \prot, of the star, which can be measured thanks to the rotational 
modulation of the star's brightness produced by an uneven distribution of cool photospheric spots. The amplitude of the photometric variations
is also a useful parameter related to the magnetic activity.

To measure \prot\ we have applied a periodogram analysis \citep{Scargle1982} and the CLEAN deconvolution algorithm \citep{Roberts1987} to the {\it TESS} light 
curves of the members of ASCC\,123.
For the stars with periods longer than a few days, we merged the data of sectors 16 and 17 to expand the time base so as to improve the precision of the period 
determination. Ultrafast rotators, such as S\,214 and S\,517, display a substantial starspot evolution during the 50-days time baseline, so 
that for these stars we have preferred to analyse independently sectors 16 and 17, obtaining always very similar values of \prot\  in the two epochs.

We retrieved the {\it TESS} light curves for the 55 candidate members of the cluster ASCC\,123 according to \citet{Cantat2018} but we were able to determine the 
rotation period only for 29 of them. Indeed, we rejected the stars with \gaia\ magnitude $G>16.5$, whose {\it TESS} photometry is too noisy for a meaningful measure
of the rotation period. Moreover, we discarded the objects that we recognized as double-lined binaries (SB2) in Paper~I. All the periods are reported in 
Table~\ref{Tab:periods}. 
For illustrative purposes, the {\it TESS} light curve and the result of the period search for S\,214 is shown in Fig.~\ref{fig:TESS_ASC214}. The light curves of the 
remaining FGK members studied in Paper~I are displayed in Figs.~\ref{fig:TESS_ASC39} to \ref{fig:TESS_F2}.
The modulation of the star brightness produced by the spots and star rotation is clearly visible for all the targets;  exceptions are the two hottest sources. 
Interestingly, the light curve of S\,39 clearly displays two dips reminiscent of a planet transit with a time separation of about 20.31\,days 
(see also Fig.~\ref{fig:TESS_Transits}). When we started the present investigation, this star was not known to host exoplanets and was not included in the list 
of {\it TESS} Objects of Interest (TOIs). 
We have then proposed this source for the {\it TESS} Follow-up Observing Program (TFOP) and we are planning ground-based observations, whose results will be 
presented in a forthcoming work. To search for the rotational period of S\,39, we excluded the two transits from the {\it TESS} light curve obtained merging 
those of sector 16 and 17 before applying the period analysis.

We note that the cleaned periodograms display for all the stars but S\,39 a clear power peak (marked with a red vertical line in each plot) without secondary
peaks of comparable amplitude. This period thus represents the rotational one. 
Even the hottest star, S\,554, displays a significant rotational modulation, although with a very low amplitude of only $\sim$\,0.007 mag, which is clearly shown 
by the phased light curve displayed in the right inset panel of Fig.~\ref{fig:TESS_ASC554}.
As mentioned above, the only doubtful case is S\,39, for which the highest power in the periodogram is found at a period 4.86\,days, which is clearly inconsistent 
with the rapid rotation of the star indicated by the high value of the projected rotation velocity, \vsini$\simeq 49$\,\kms. However, the periodogram displays other 
peaks at about 7 days (which is still more inconsistent with the \vsini) and at 1.67 days (marked with a blue vertical line in the left inset panel of 
Fig.\,\ref{fig:TESS_ASC39}); we consider the last as a possible \prot\ value and list it in Table~\ref{Tab:param} with a colon.

\subsection{Chromospheric emission}
\label{Sec:chrom}

\begin{figure*}
\includegraphics[width=9.2cm]{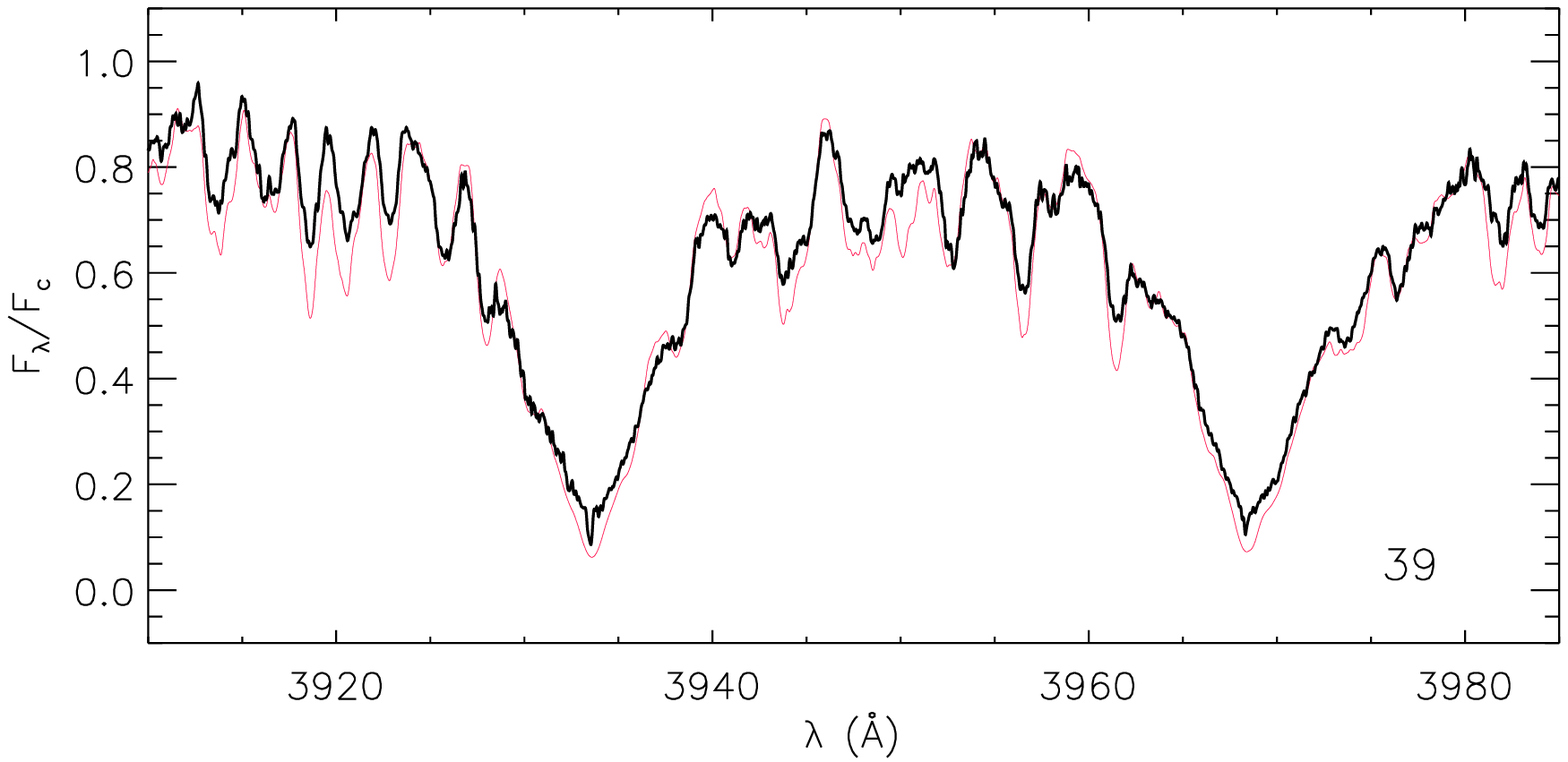}       
\hspace{-.8cm}
\includegraphics[width=9.2cm]{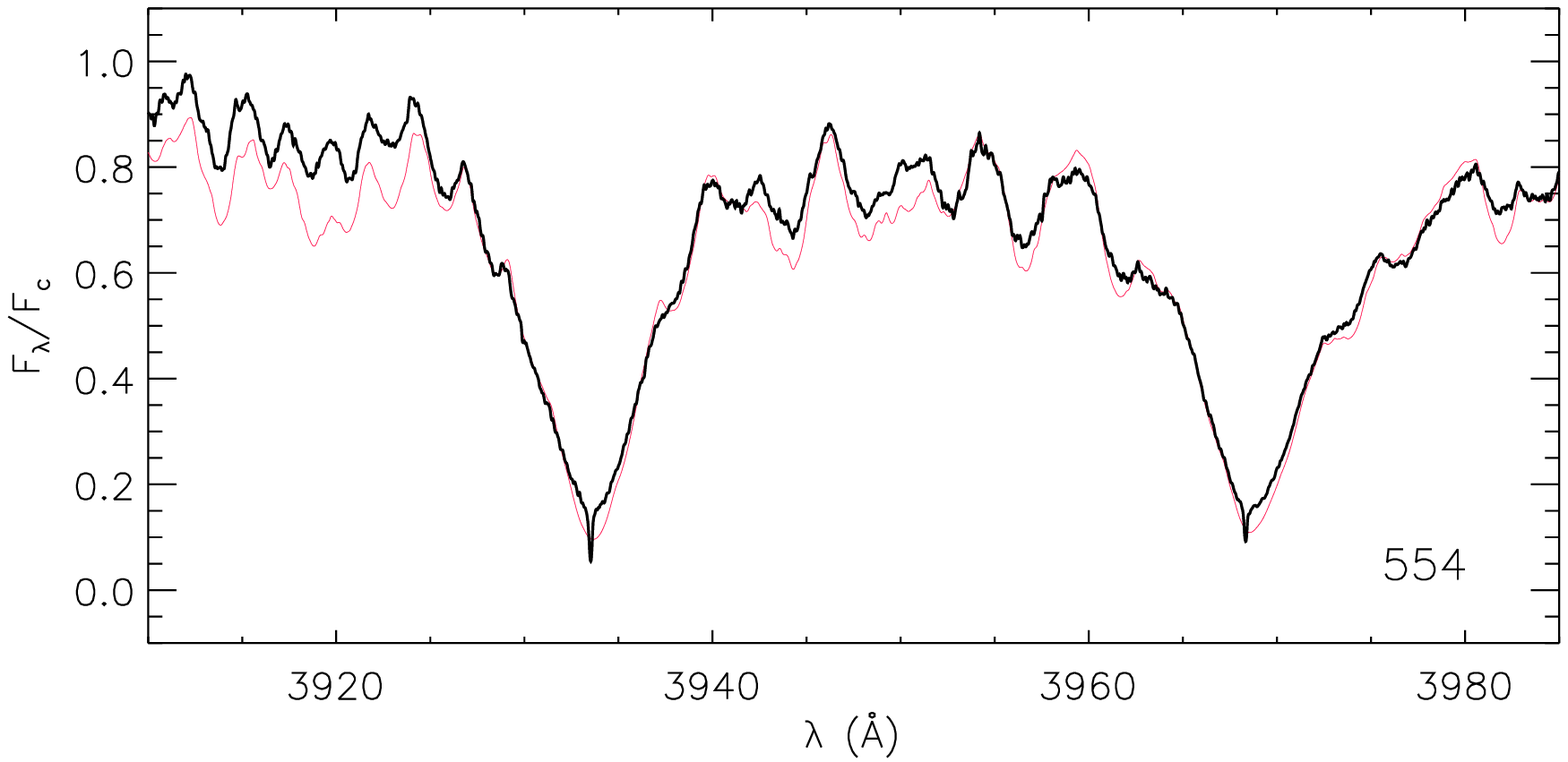}      
\hspace{-.8cm}
\includegraphics[width=9.2cm]{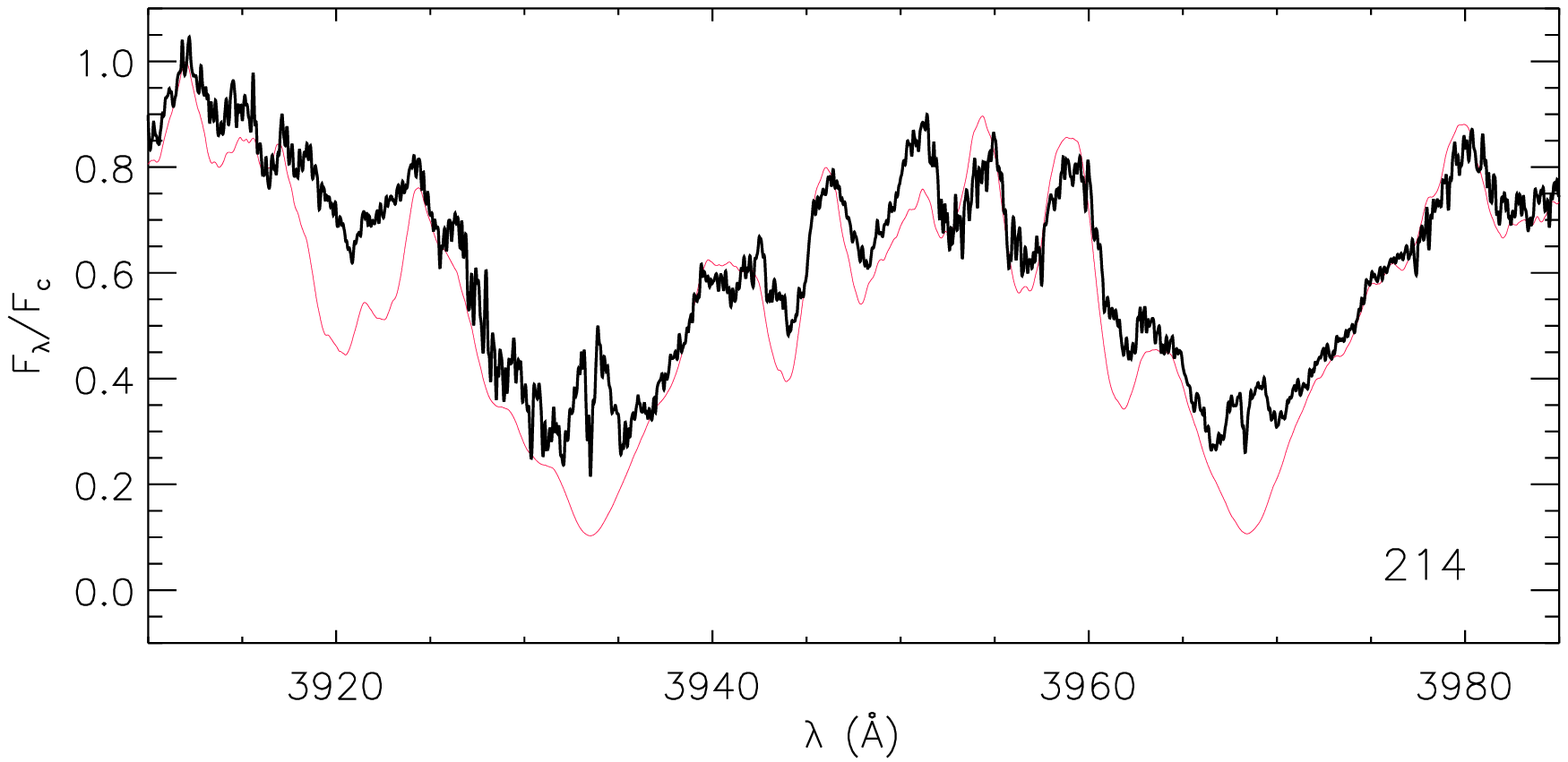}      
\hspace{-.8cm}
\includegraphics[width=9.2cm]{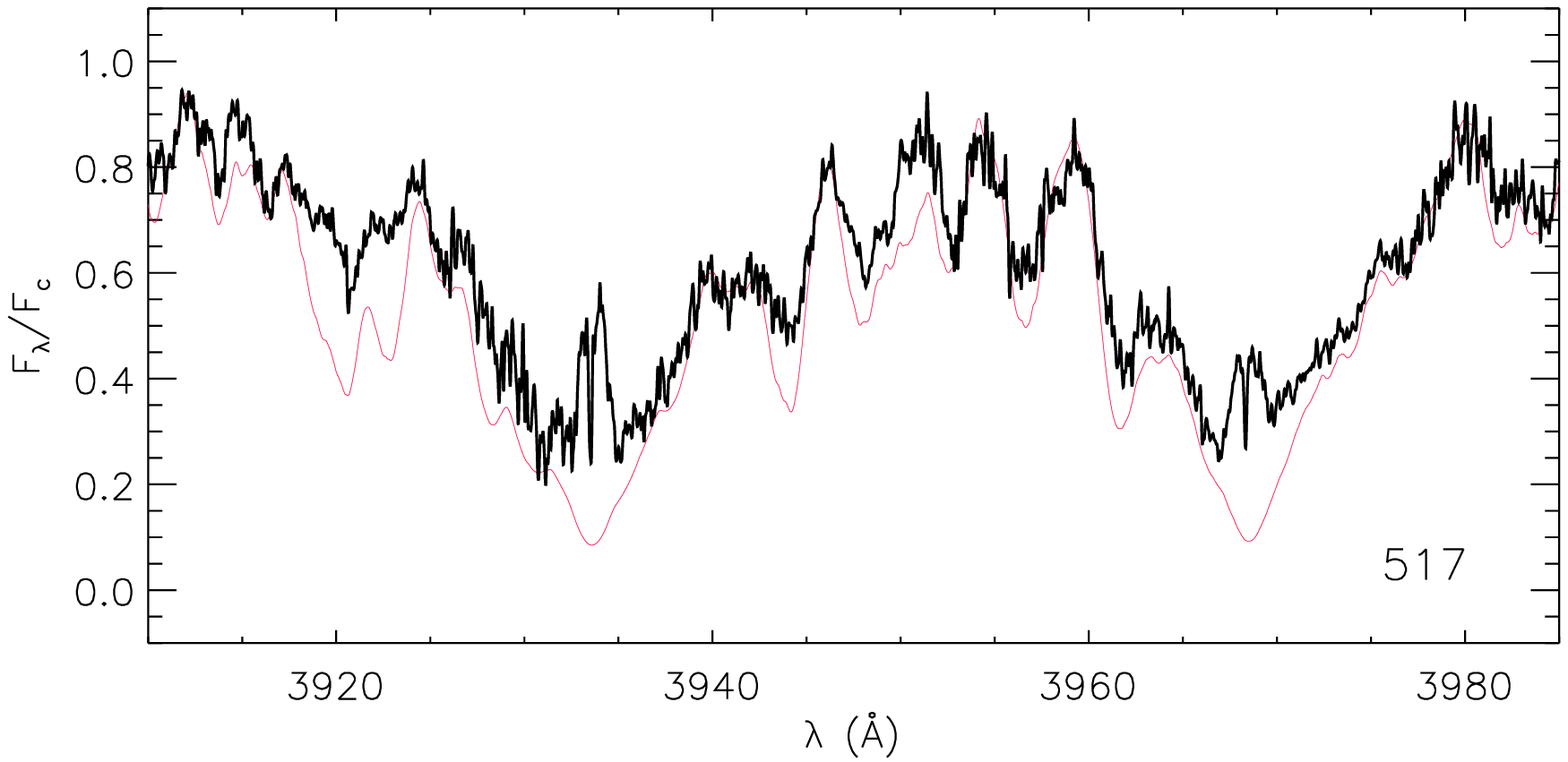}      
\hspace{-.8cm}
\includegraphics[width=9.2cm]{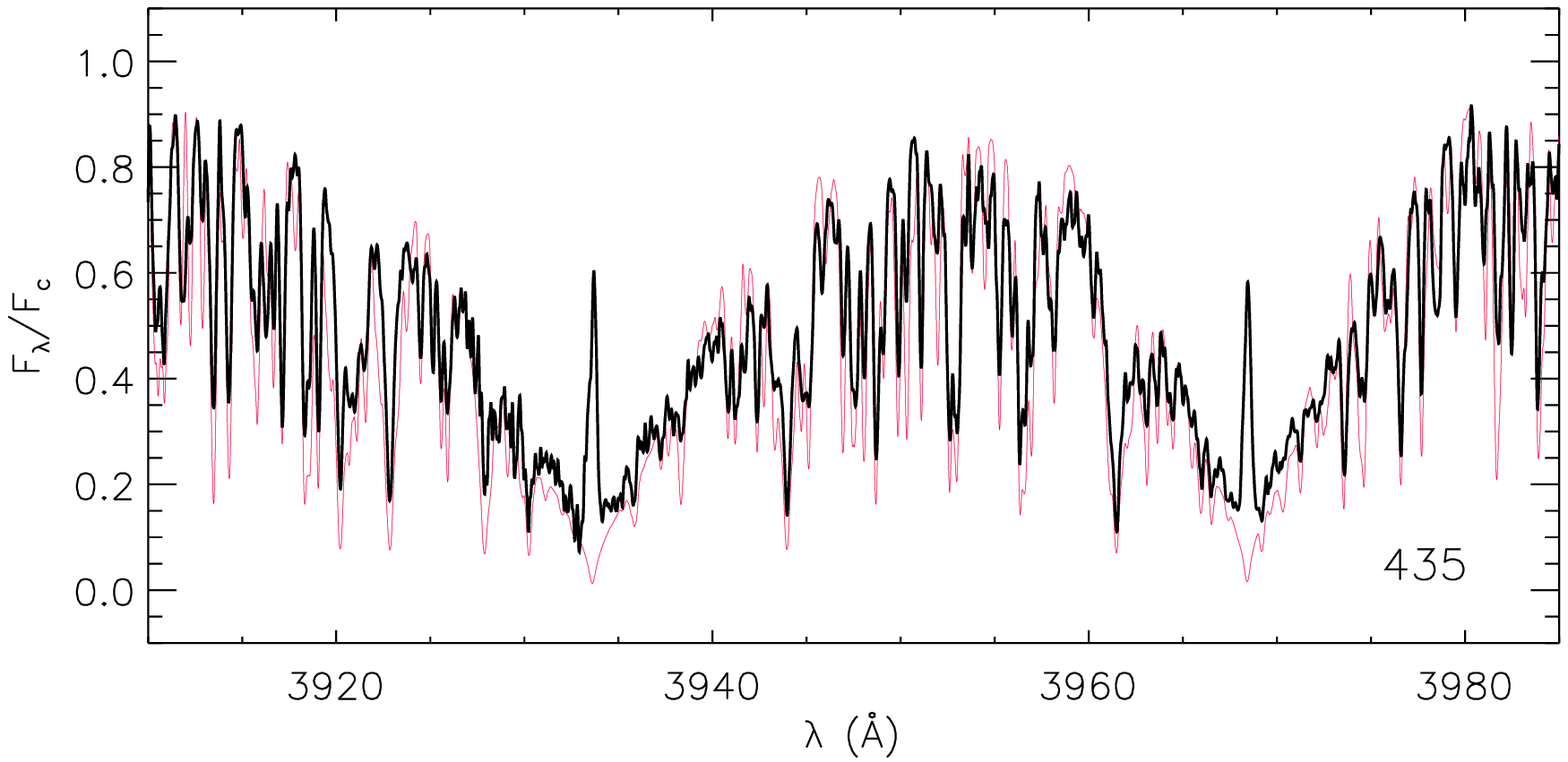}      
\hspace{-.8cm}
\includegraphics[width=9.2cm]{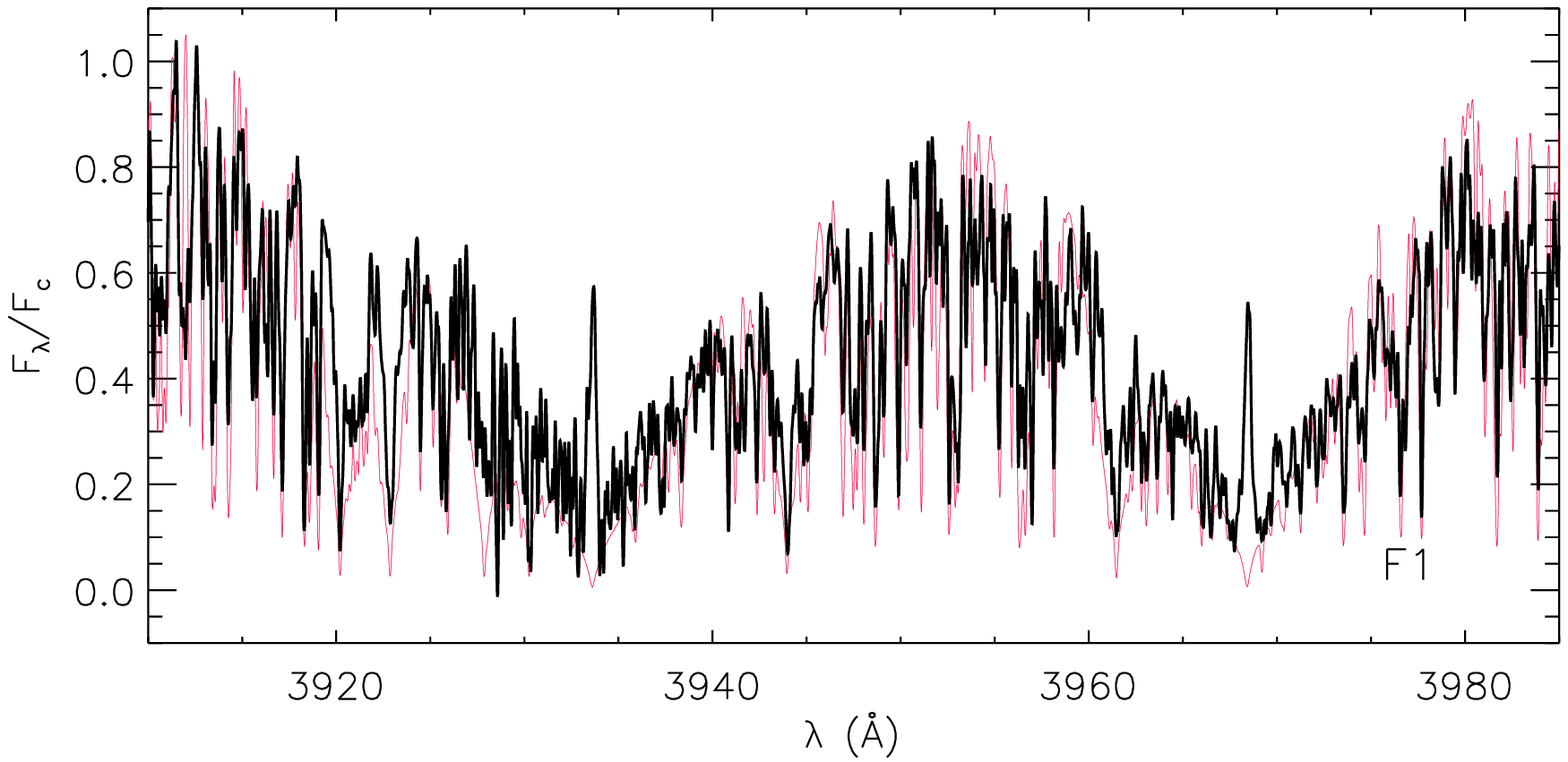}	
\hspace{-.8cm}
\includegraphics[width=9.2cm]{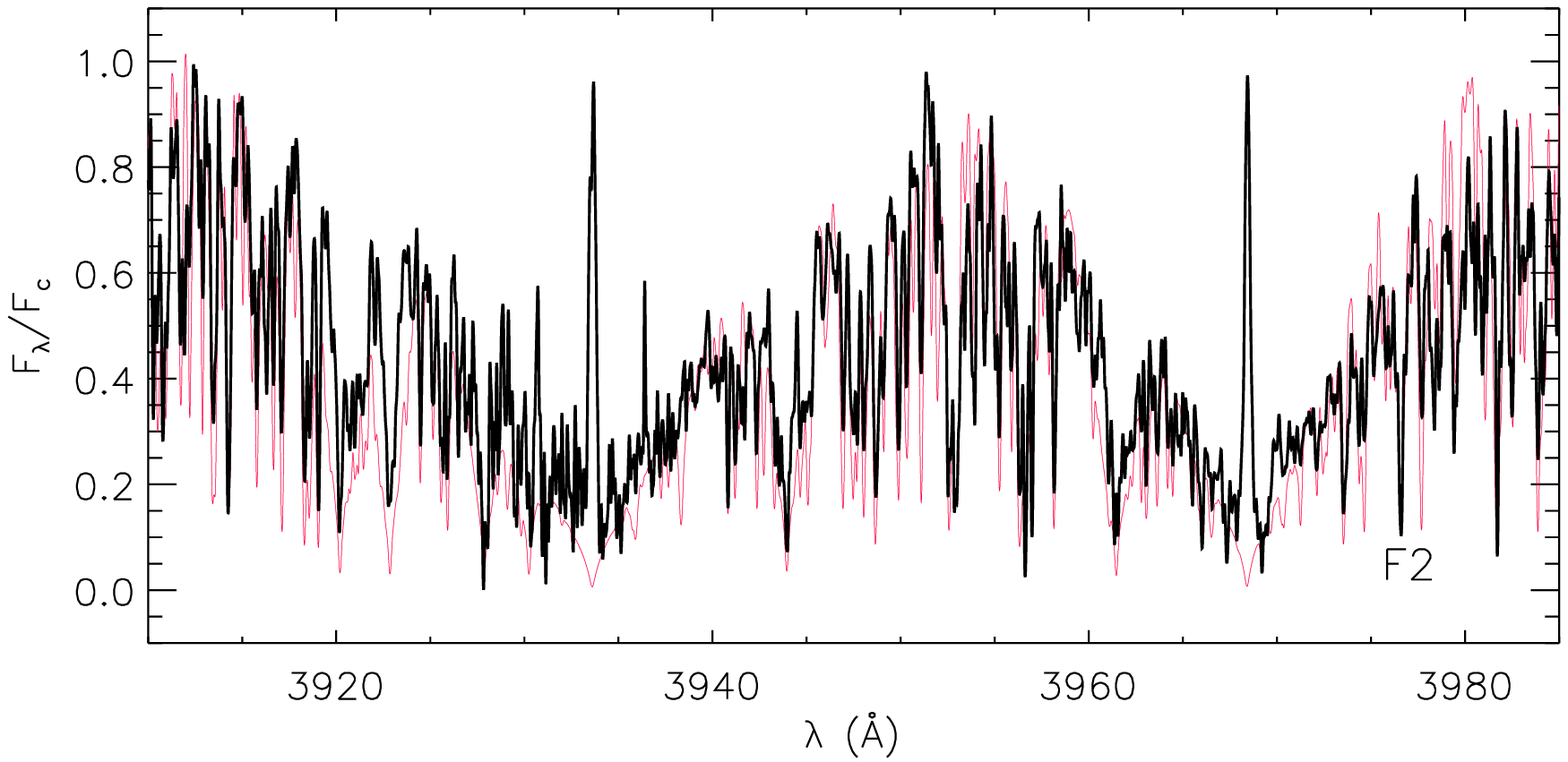}	
\vspace{0cm}
\caption{HARPS-N spectra of the investigated stars in the \ion{Ca}{ii} H\&K region. 
In each box, the non-active template (red line) is overlaid with the observed spectra (thick black line). 
The ID of the source is marked in the lower right corner of each box. The two uppermost panels display the hottest 
(F4V-type) stars in our sample for which no clear emission in the H\&K line cores is visible. }
\label{fig:subtraction_CaII}
\end{figure*}

The level of chromospheric activity can be evaluated from the emission in the core of the H$\alpha$ and \ion{Ca}{ii} H and K lines
\citep[see, e.g.,][and references therein]{Frasca1994,Frasca2010,Frasca2019}.  
To this end, we subtracted photospheric templates from the observed spectra of the targets, 
to remove the underlying photospheric absorption lines so as to leave as residual the chromospheric emission in the cores of H$\alpha$ 
and  \ion{Ca}{ii} lines. The photospheric templates are made with synthetic BTSettl spectra \citep{Allard2012}, for \ion{Ca}{ii} lines, 
and spectra of non-active stars for the H$\alpha$; indeed the latter are better reproducing the core of the H$\alpha$ line in absence of a 
significant chromospheric contribution. 
Since the signal-to-noise ratio in  the \ion{Ca}{ii} H\,\&\,K region is quite low, ranging from 3 (for the coolest members) to 38 
(for the hottest one), we have degraded the original resolution  (R=115,000) of the HARPS-N spectra and the synthetic spectra to R=42,000, 
which is the same of the ELODIE templates used for the determination of the atmospheric parameters 
and the subtraction of the photospheric H$\alpha$ template (Paper~I).
Furthermore, the photospheric templates have been aligned in wavelength with the target spectra by means of the cross-correlation function 
and have been rotationally broadened by the convolution with a rotational profile with the \vsini\ of the target star (Table~\ref{Tab:param}).

The observed spectra (black lines) and the photospheric templates (red lines) are displayed in Fig.~\ref{fig:subtraction_CaII} for the 
\ion{Ca}{ii} H\,\&\,K region.
Figure~\ref{fig:subtraction_halpha} shows the H$\alpha$ profiles of the targets (black dots) overlaid to the photspheric templates (red lines)
and the difference of the two (blue lines).
The excess H$\alpha$ equivalent width, $W_{\rm H\alpha}$, has been obtained by integrating the residual H$\alpha$ emission profile, which
is highlighted by the green hatched areas in the difference spectra in each box of Fig.~\ref{fig:subtraction_halpha}.
Analogously, for each target, the difference between observed and template spectrum in the \ion{Ca}{ii} region leaves emission excesses in the 
cores of the H and K lines, which have been integrated obtaining the equivalent widths, $W_{\rm CaII-H}$ and $W_{\rm CaII-K}$. 
It is worth noticing that the hottest star in our sample, S\,554 (\teff=6871\,K), does not show any filling in the H$\alpha$ core and only
a tiny filling in the cores of the \ion{Ca}{ii} lines (upper right panels in Figs.~\ref{fig:subtraction_halpha} and
\ref{fig:subtraction_CaII}). For S\,39 (\teff=6667\,K) a small residual emission is detected in the core of the H$\alpha$ and also 
in the \ion{Ca}{ii} lines.  However, unlike the other stars of lower temperature, no reversal in emission emerges in the cores of the H and K lines
for these two F4-type stars. This is in line with what is expected on the basis of the reduced chromospheric activity in these stars with shallow 
convective envelopes and the difficulty of detecting low chromospheric emission against a large continuum flux.

As a more effective indicator of chromospheric activity, we computed the surface flux in each of the chromospheric lines, $F_{\rm line}$.
For the \ion{Ca}{ii}\,K line this reads as

\begin{equation}
F_{\rm CaII-K}  =  F_{3933}W_{\rm CaII-K}, 
\end{equation}

{\noindent where $F_{3933}$ is the flux at the continuum at the center of the \ion{Ca}{ii} K line per unit stellar surface area, which is evaluated 
from the BTSettl spectra \citep{Allard2012} at the stellar temperature and surface gravity of the target. We evaluated the flux error by taking 
into account the error of $W_{\rm CaII-K}$ and the uncertainty in the continuum flux at the line center, $F_{3933}$, which is estimated 
considering the errors of \teff\  and \logg. }

We computed surface fluxes for the other chromospheric diagnostics in the same way as for the \ion{Ca}{ii}-K line. The EWs and fluxes are reported in 
Table~\ref{Tab:Halpha_CaII}.

\setlength{\tabcolsep}{4pt}

\begin{table*}
\caption{H$\alpha$ and \ion{Ca}{ii} H\,\&\,K equivalent widths and fluxes.}
\begin{center}
\begin{tabular}{lrrrccrrrrcccc}
\hline
\noalign{\smallskip}
ID      & \teff &  $W_{\rm H\alpha}$    & err & $F_{\rm H\alpha}$ & err & $W_{\rm CaII-K}$ & err & $W_{\rm CaII-H}$  & err & $F_{\rm CaII-K}$ & err  & $F_{\rm CaII-H}$ & err \\   
        & (K)   &  \multicolumn{2}{c}{(m\AA)} & \multicolumn{2}{c}{(erg\,cm$^{-2}$s$^{-1}$)} & \multicolumn{2}{c}{(m\AA)} & \multicolumn{2}{c}{(m\AA)}  & \multicolumn{2}{c}{(erg\,cm$^{-2}$s$^{-1}$)} & \multicolumn{2}{c}{(erg\,cm$^{-2}$s$^{-1}$)}  \\  
\hline
\noalign{\smallskip}
  39    &  6667 &   73  &   13  & 9.01e+05 & 1.69e+05  &  189 &  65  & 145 &   55  & 3.42e+06  & 1.23e+06  & 2.62e+06 &  1.05e+06  \\ 
 214    &  5804 &  291  &   21  & 2.17e+06 & 1.99e+05  &  853 & 122  & 780 &  127  & 5.03e+06  & 1.04e+06  & 4.60e+06 &  1.02e+06  \\ 
 435    &  5758 &  144  &   18  & 1.04e+06 & 1.40e+05  &  492 &  74  & 438 &   80  & 2.73e+06  & 5.17e+05  & 2.43e+06 &  5.27e+05  \\ 
 517    &  5784 &  591  &   33  & 4.34e+06 & 3.31e+05  & 1082 & 211  & 985 &  232  & 6.18e+06  & 1.40e+06  & 5.63e+06 &  1.48e+06  \\ 
 554    &  6871 & \dots & \dots & \dots    & \dots     &   64 &  22  &  64 &   20  & 1.45e+06  & 5.32e+05  & 1.43e+06 &  4.83e+05  \\ 
  F1    &  5263 &  196  &   22  & 9.93e+05 & 1.33e+05  &  519 & 155  & 443 &  174  & 1.34e+06  & 4.57e+05  & 1.15e+06 &  4.87e+05  \\ 
  F2    &  5237 &  396  &   30  & 1.96e+06 & 1.93e+05  &  786 & 194  & 608 &  250  & 1.94e+06  & 5.54e+05  & 1.50e+06 &  6.55e+05  \\ 
\hline
\end{tabular}
\end{center}
\label{Tab:Halpha_CaII}
\end{table*}

\section{Discussion}
\label{Sec:discussion}

\subsection{Rotational periods}
\begin{figure}
\begin{center}
\includegraphics[width=9.cm]{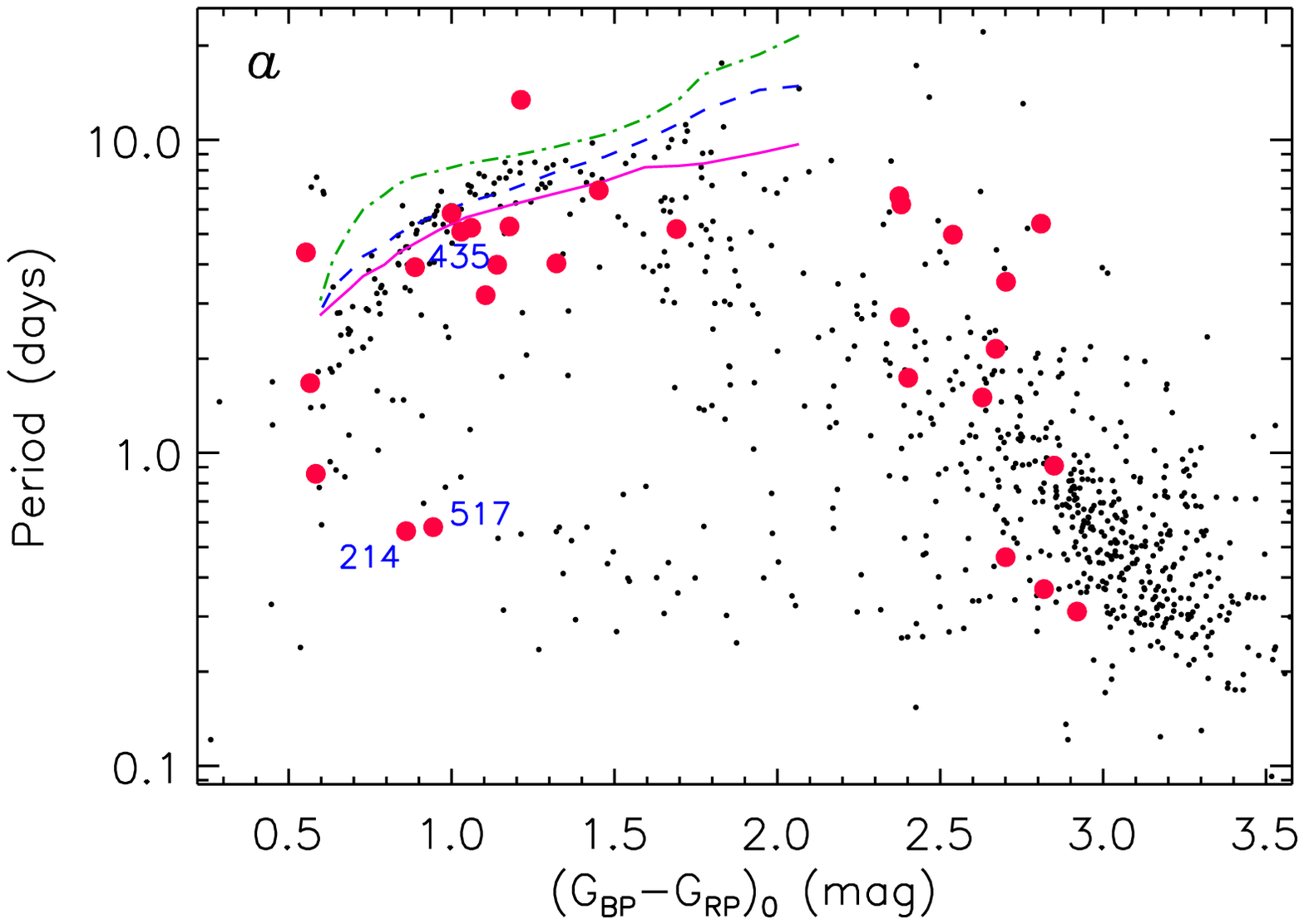}	
\includegraphics[width=9.cm]{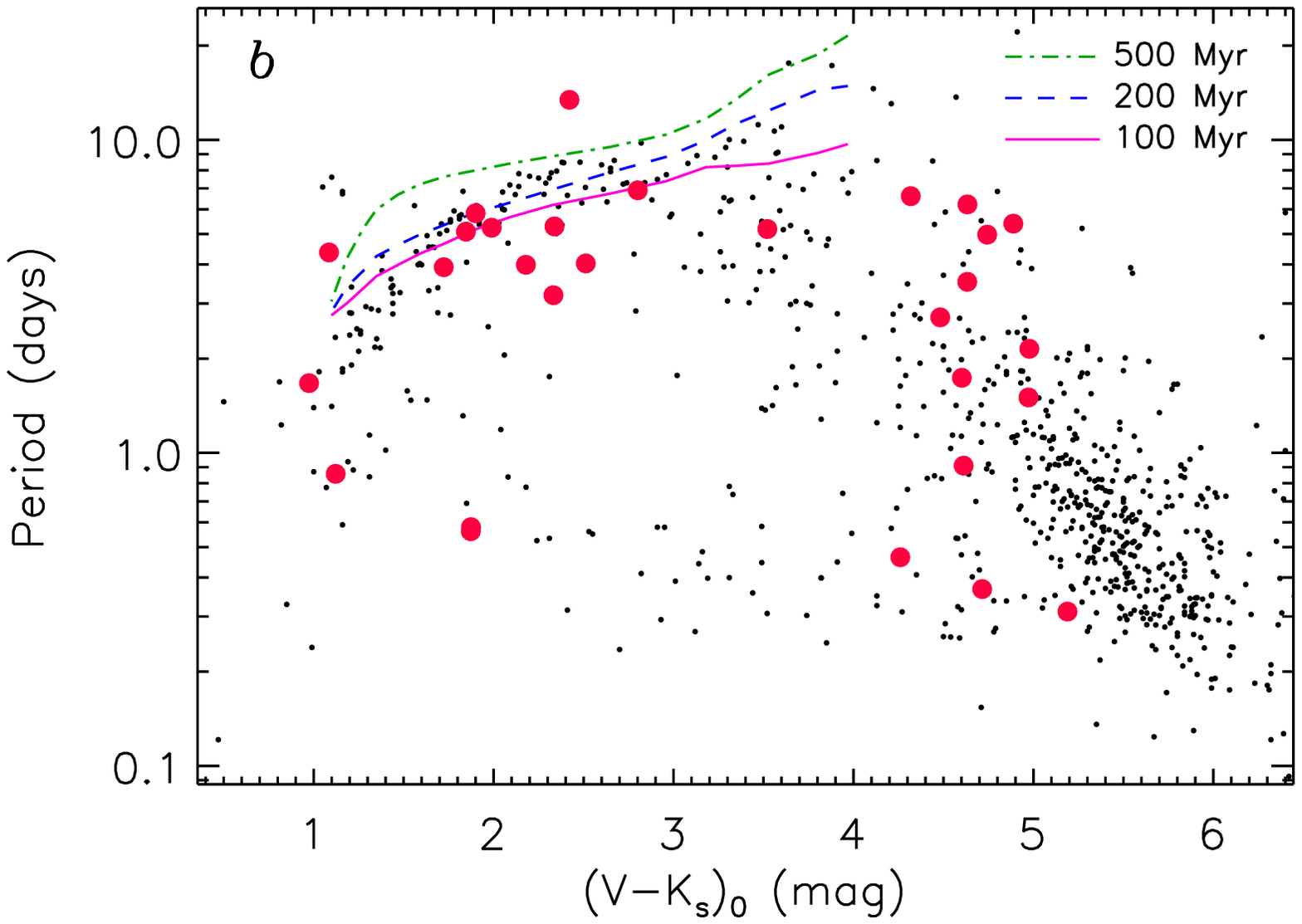}	
\vspace{-0.3cm}
\caption{{\it (a)} Rotation periods versus the dereddened colour index $(G_{\rm BP}-G_{\rm RP})_0$ for the candidate members of the cluster ASCC\,123 selected by 
\citet{Cantat2018} (red circles). Small black dots denote the periods derived for a sample of Pleiades members by \citet{Rebull2016} during the {\it Kepler}-K2 campaign. 
The ID of the three solar-like stars is also marked. Rotational isochrones from \citet{Spada2020} at ages of 100, 200, and 500 Myr are over-plotted with continuous, 
dashed, and dash-dotted lines, respectively. 
{\it (b)} Rotation periods versus $(V-K_{\rm s})_0$.  }
\label{fig:Prot}
\end{center}
\end{figure}

The distribution of rotation periods as a function of the stellar mass is a very useful tool for characterising the evolutionary stage of a cluster 
\citep[e.g.][and references therein]{Barnes2003}.
To this aim, we plot in Fig.~\ref{fig:Prot}a the periods listed in Table~\ref{Tab:periods} versus the colour index $(G_{\rm BP}-G_{\rm RP})_0$, which 
has been dereddened using the value $A_V=0.13$\,mag derived in Paper~I and the extinction relations in \citet{WangChen2019}. In the same plot we also 
show the periods derived by \citet{Rebull2016} for a sample of Pleiades members observed with {\it Kepler}-K2. 
To compare our period distribution with previous works devoted to other clusters we also use the colour index $(V-K_{\rm s})_0$ in Fig.~\ref{fig:Prot}b.
Despite the small data sample, the rotation periods for the candidate members of ASCC\,123 
roughly follow the distribution of the Pleiades, with the presence of both fast and slow rotators for G-type stars.
We note that, among the three solar-like stars, S\,214 and S\,517 lie in the fast-rotating {\it C} sequence (according to the notation of \citealt{Barnes2003}), 
while S\,435 falls on the {\it I} sequence of the Pleiades, for which 
the period increases steadily with the colour index up to $(G_{\rm BP}-G_{\rm RP})_0\approx 1.5$\,mag and then decreases reaching the {\it C} sequence at 
$(G_{\rm BP}-G_{\rm RP})_0\approx 3$\,mag. 
We do not find fast rotators (in the lower {\it C} sequence) with colour 
$1<(G_{\rm BP}-G_{\rm RP})_0<2.5$, i.e. approximately in the spectral-type range G3--M3. However, the presence of the two fast-rotating solar-type stars strongly 
supports a young age for the cluster, because the {\it C} sequence of older clusters like NGC\,3532 (age$\sim$300\,Myr)  displaces towards cooler objects 
\citep[e.g.,][]{Barnes2003,Fritzewski2021} and practically disappears for the still older Hyades or Praesepe clusters \citep{Barnes2003,Rebull2022}, i.e. at
age $\approx$\,600\,Myr. Indeed, the hottest stars in NGC\,3532 with $P_{\rm rot}<1$\,day have a $(V-K_{\rm s})_0\simeq 3$ \citep{Fritzewski2021}. Another cluster 
in the age range between the Pleiades and the Hyades is M\,48. For this cluster, with an age of about 450\,Myr, \citet{Barnes2015} found no G- or early K-type fast 
rotating stars. 
A younger cluster showing a clear {\it C} sequence reaching the G-type stars is M\,34 \citep[age$\sim$250\,Myr,][]{Ianna1993}. Therefore, the gyrochronogical age 
of ASCC\,123 should not be greater than 250--300\,Myr\footnote{However, recent age determinations based on the {\it Gaia} CMDs point to a younger age of about 
130\,Myr for M\,34 \citep[see, e.g.,][]{Bossini2019,Cantat2020}}. 

As a further check, we have also plotted in Fig.\,\ref{fig:Prot} rotational isochrones calculated for slow-rotator sequences by \citet{Spada2020} and expressed as a 
function of stellar mass or $B-V$ in their Table\,A.1. For this purpose, we have converted the $B-V$ colour index into $V-K_S$ and $G_{\rm BP}-G_{\rm RP}$ according 
to the calibrations of Mamajek\footnote{\url{http://www.pas.rochester.edu/~emamajek/EEM_dwarf_UBVIJHK_colors_Teff.txt}}. 
As apparent, the $I$ sequences of both ASCC\,123 and the Pleiades agree with the theoretical isochrone of \citet{Spada2020} with the smallest age in their model 
($\tau=100$\,Myr).

On the other hand, a much younger age seems to be ruled out by the clear gap between fast and slow rotators observed for ASCC\,123. Indeed, this gap is not visible 
in very young ($\approx 16$\,Myr) stellar populations, such as the Upper Centaurus-Lupus (UCL) and Lower Centaurus-Crux (LCC) \citep{Rebull2022}.
The most discrepant object, with respect to the Pleiades diagram, is TIC\,420123691, for which we measure a period of 13.4 days that places it much higher than the Pleiades {\it I} sequence. This object could be a non-member or a close binary.

\subsection{Variation amplitudes}

\begin{figure}
\begin{center}
\includegraphics[width=9.cm]{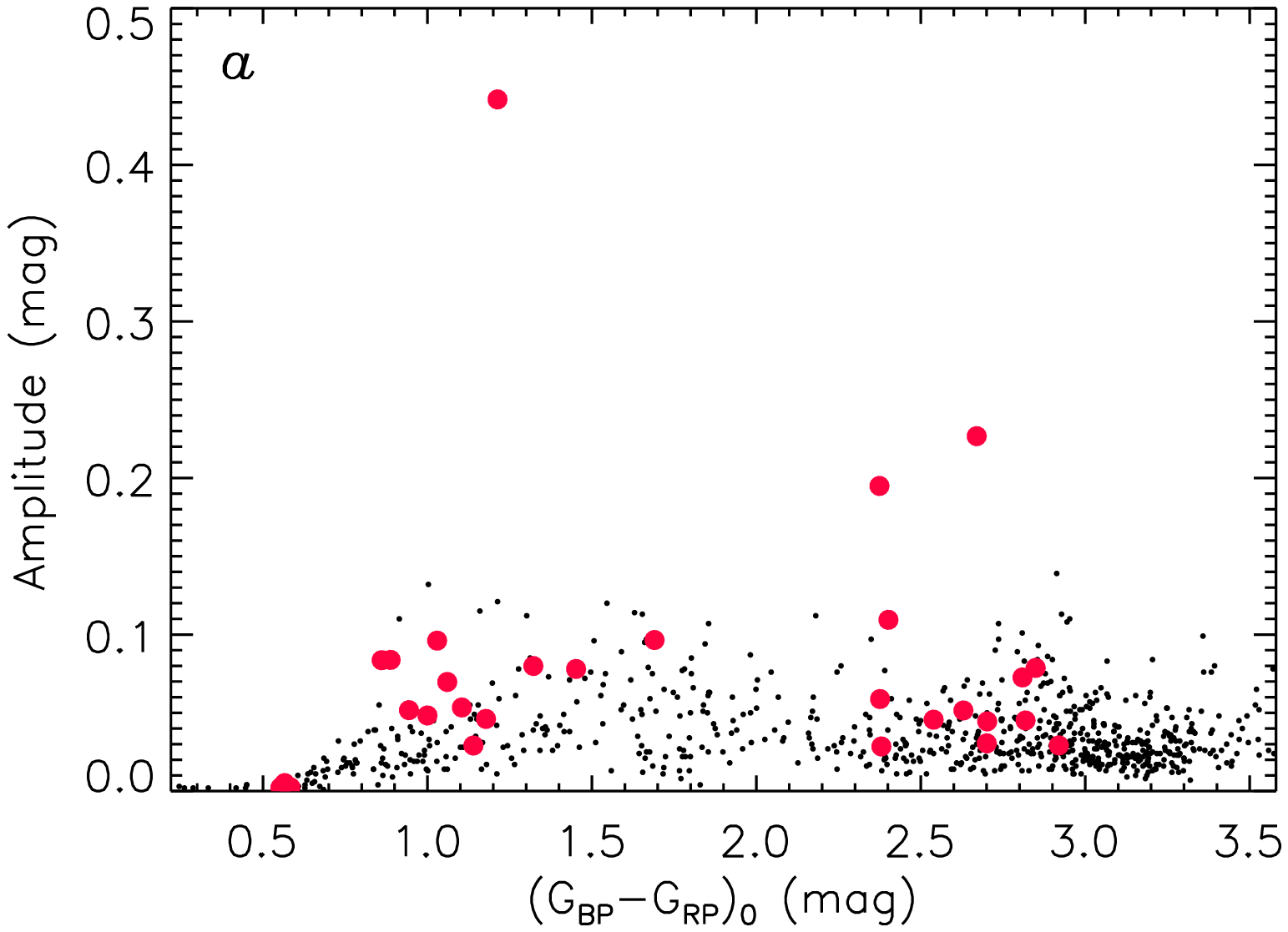}	
\includegraphics[width=9.cm]{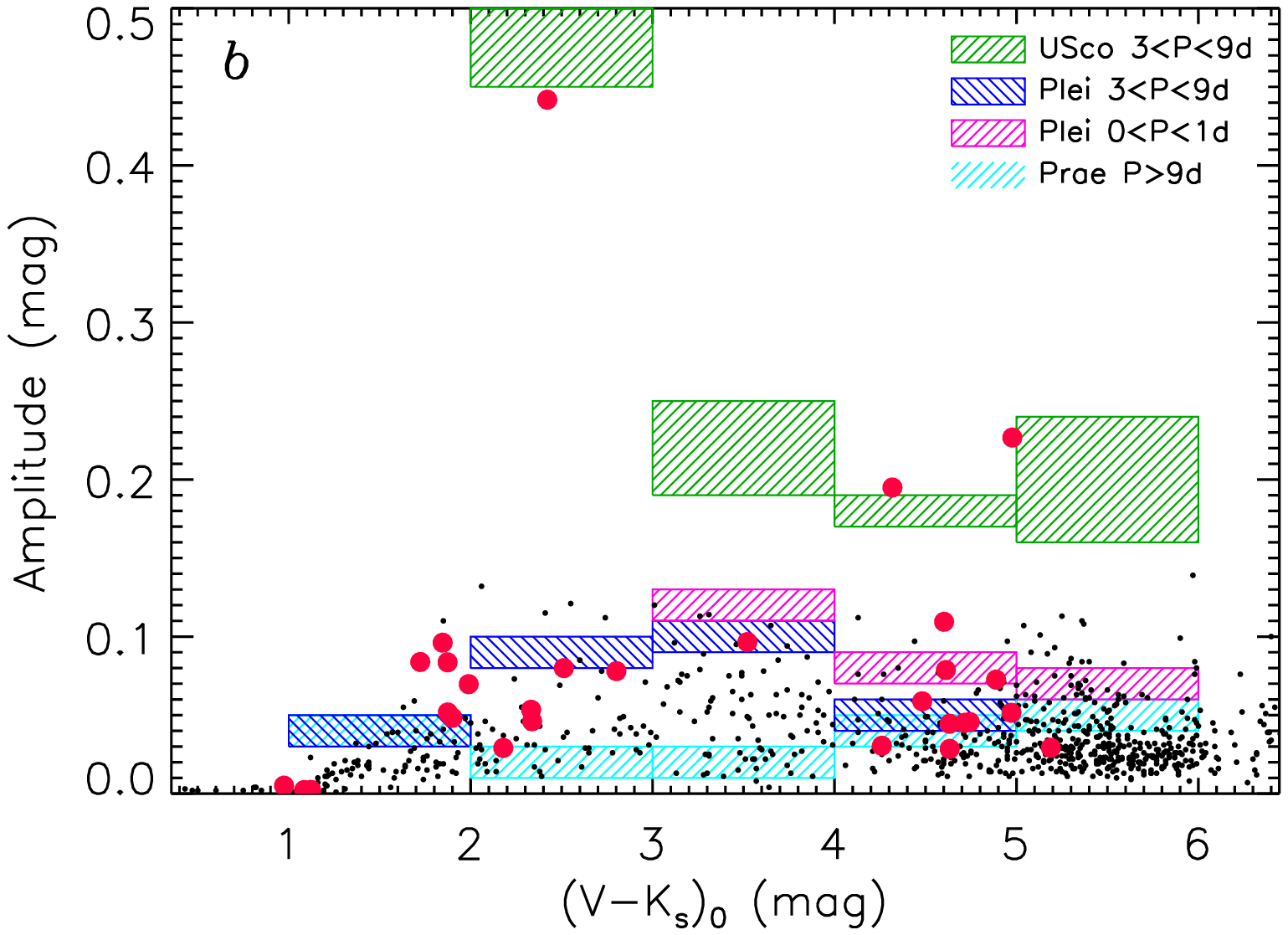}	
\vspace{-0.3cm}
\caption{{\bf a)} Variation amplitudes versus the dereddened colour index $(G_{\rm BP}-G_{\rm RP})_0$ for the candidate members of the cluster ASCC\,123  
(red circles) and for the sample of Pleiades members (small black dots) studied by \citet{Rebull2016}. 
{\bf b)} Amplitudes as a function of the $(V-K_{\rm s})_0$ colour index. The meaning of the symbols is as in panel (a). The hatched rectangles represent 
the amplitudes at the 80th percentile for the clusters and the period ranges indicated in the legend according to \citet{Messina2021}.
}
\label{fig:Ampl}
\end{center}
\end{figure}

Another proxy for the magnetic activity level is the amplitude of photometric variations induced by starspots \citep[see, e.g.][]{See2021, Messina2021}.
The variation amplitudes of our sources, which are reported in Table~\ref{Tab:periods}, were measured on their {\it TESS} light curves, converting fluxes into 
magnitudes and rejecting 5-$\sigma$ outliers. We excluded long-term linear trends by taking suitable data chunks and took the largest variation amplitude.
We show the amplitude of variation as a function of the $(G_{\rm BP}-G_{\rm RP})_0$ colour index in Fig.~\ref{fig:Ampl}a, where the same data for the Pleiades 
\citep{Rebull2016} are also plotted for comparison.
TIC\,420123691, with an amplitude of 0.44\,mag is the most discrepant object also in this diagram. A spectroscopic follow-up would be very useful to better 
clarify its nature. A similar plot with the $(V-K_{\rm s})_0$ colour in the x axis is shown in Fig.~\ref{fig:Ampl}b, where we overlay the amplitudes at the 80th 
percentile of three clusters (Upper Sco, Pleiades, and Praesepe) in bins of the colour index as derived by \citet[][see his Table~1]{Messina2021}. 
For the Pleiades, we plot the 80th-percentile amplitudes of both the fast-rotating ($0.1<P_{\rm rot}<1.0$~days) and slow-rotating stars ($3<P_{\rm rot}<9$~days), 
while for the Praesepe we do not have the sequence of fast rotators in this colour range, so we plot only the amplitude of slow rotators ($P_{\rm rot}>9$~days). 
Indeed, for the Praesepe stars with $P_{\rm rot}>9$~days, \citet{Messina2021} reports the 80th-percentile data for all the bins of $(V-K_{\rm s})_0$ in the 
range 1--6 mag. However, we note that the amplitudes of the Praesepe stars with $3<P_{\rm rot}<9$~days for the three $(V-K_{\rm s})_0$ bins for which they could be 
determined are basically the same as those in the longest period range.
The comparison of the variation amplitudes with the loci of the three clusters further indicates that ASCC\,123 is clearly older than Upper Sco 
(age\,$\approx$\,10\,Myr, \citealt{Feiden2016}) and younger than Praesepe, suggesting an age similar to that of the Pleiades.

\begin{figure}
\begin{center}
\includegraphics[width=9.cm]{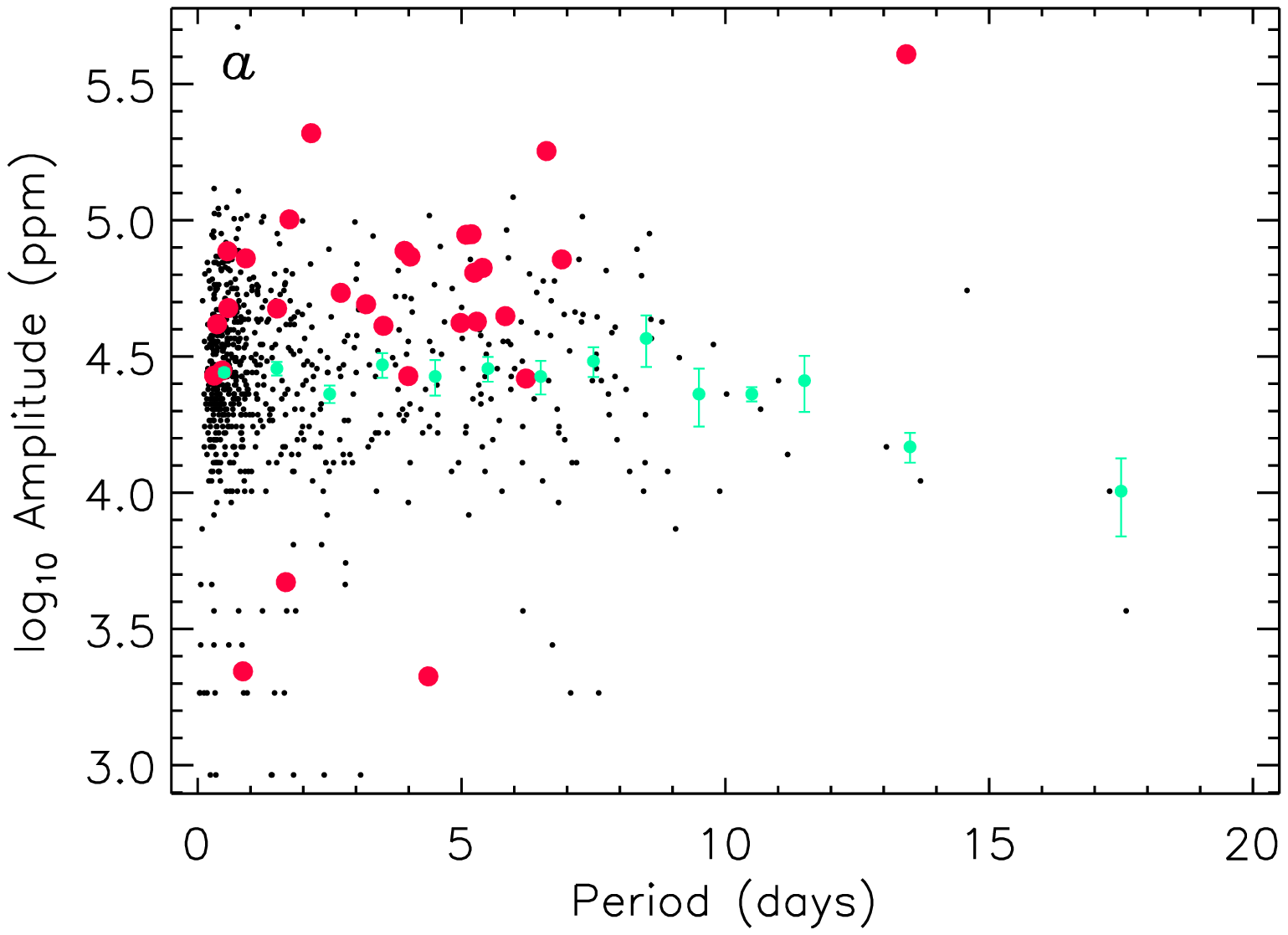}
\includegraphics[width=9.cm]{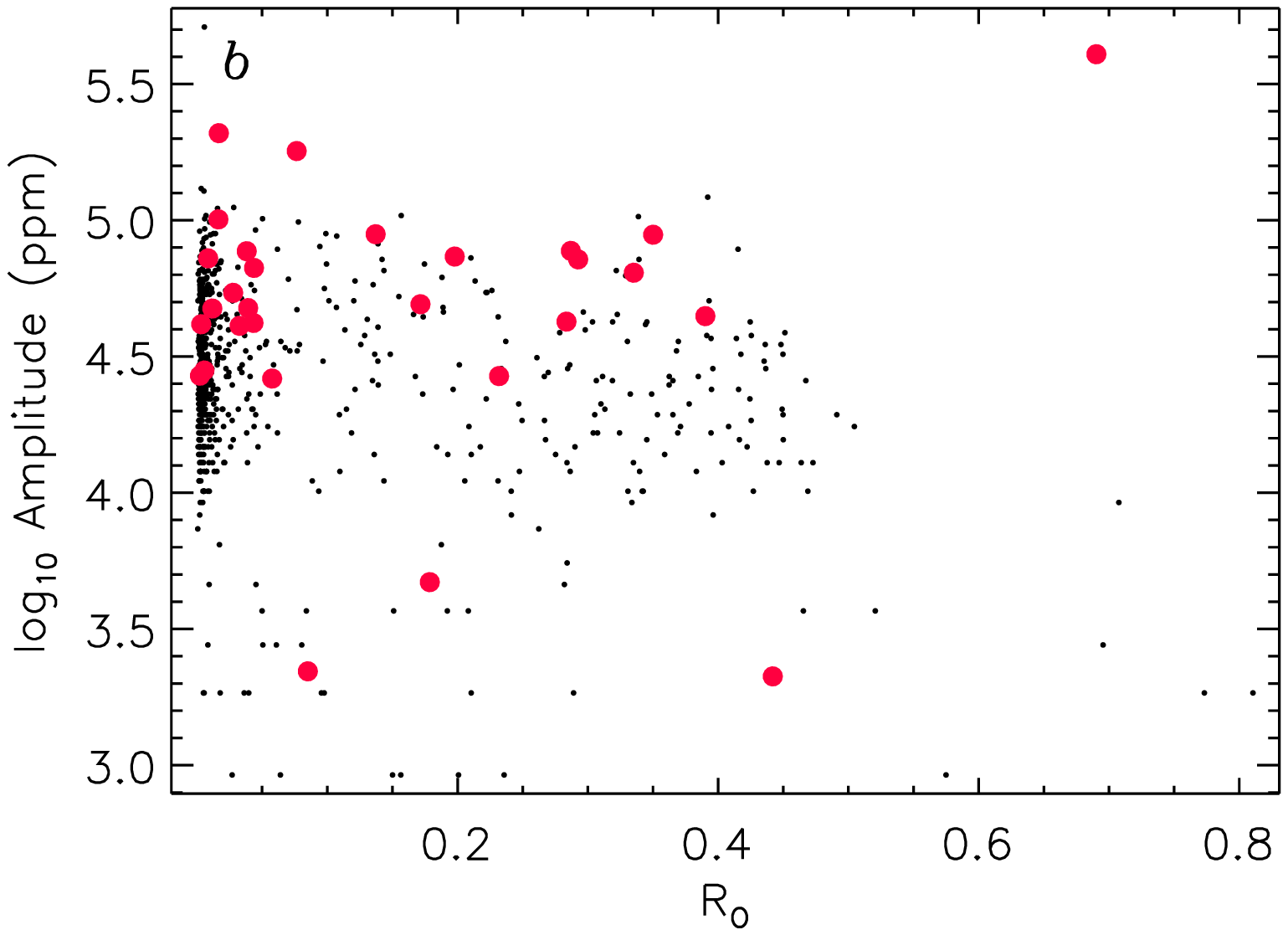}
\vspace{-0.3cm}
\caption{{\bf a)} Variation amplitudes versus the rotation period for the candidate members of the cluster ASCC\,123 (red circles).
Small black dots denote the amplitudes for the Pleiades members by \citet{Rebull2016}. The green dots and errorbars show the median variability range for the Pleiades for period bins of one day and the standard deviation therein.
{\bf b)} Variation amplitudes versus the Rossby number $R_{\rm O}=P_{\rm rot}/\tau_{\rm c}$.
}
\label{fig:Ampl_Prot}
\end{center}
\end{figure}

One of the main ingredients of dynamo mechanisms generating magnetic fields and operating in stellar interiors is the rotation rate. Therefore, a correlation of 
magnetic field intensity (or proxies of magnetic field) and the rotation period is expected. Indeed, the correlation with \prot\ has been widely documented in the 
literature for several diagnostics of magnetic activity at chromospheric and coronal levels \citep[see, e.g.,][]{Cardini2007,Frasca2016,Pizzolato2003,Reiners2014}. 
As regards the photospheric activity, traced, e.g., by the variation amplitude, correlation with \prot\ has been found by \citet{Reinhold2020}, who subdivided their 
very large sample of stars with periods and amplitudes detected from $Kepler$-K2 light curves in several subsamples with 200\,K temperature bins. They found a general 
decrease of variability with increasing rotation period for all temperature bins, although the data show large scatter. We plot the variation amplitude as a function 
of \prot\ in Fig.~\ref{fig:Ampl_Prot}a. No clear decrease of the amplitude with increasing \prot\ is visible either for ASCC\,123 or the Pleiades. This result is 
supported by the low Spearman's correlation coefficient, both for the Pleiades ($\rho\simeq -0.02$) and ASCC\,123 ($\rho\simeq+0.12$, even excluding the discrepant 
star with $P_{\rm rot}=13.4$\,days). However, we note that our data include stars with very different \teff\ values (and consequently different internal structure). 
As pointed out in several works, including some of those mentioned above, a better variable related to the dynamo action is the Rossby number, 
$R_{\rm O}=P_{\rm rot}/\tau_{\rm c}$, defined as the ratio of the rotation period and the convective turnover time $\tau_{\rm c}$. 
The latter is not a directly measurable variable, but can be derived from theoretical models for MS stars or from calibrations as a function of temperature or colour 
indices. In our case, we have used the empirical relation proposed by \citet[][Eq.\,10]{Wright2011} as a function of $V-K_S$.  Some correlation, albeit with large 
scatter, is found between amplitude and $R_{\rm O}$ (Fig.\,\ref{fig:Ampl_Prot}b), with Spearman's coefficients $\rho\simeq-0.16$ and $-0.13$ for ASCC\,123 and the 
Pleiades, respectively.

\subsection{Stellar spins orientation}
\label{Subsec:stellar_spins}
For the stars studied in Paper~I, for which we measured accurate values of \vsini, and thanks to the rotational periods, \prot, and 
stellar radii, $R_*$, which are reported in Table~\ref{Tab:param}, we can derive the inclination of the rotation axis as:

\begin{equation}
\sin i = \frac{v\sin i}{2\pi R_*}P_{\rm rot}.
\label{Eq:sini}
\end{equation}

{\noindent We found a value of $\sin i<1$ for all the stars, with the exception of S\,39. For it, the peak at 1.67\,days is marginally consistent (considering the errors) 
with the value of \vsini\,$\simeq\,$49\,\kms\ measured in Paper~I, implying an inclination close to 90\degr, but the {\it TESS} light curve phased with this period 
does not show any clear rotational modulation. We therefore consider this period an uncertain value. However, if the spin axis is nearly parallel to the orbital 
axis of the presumed transiting object, the inclination of the rotation axis should be very close to 90\degr, making 1.67 days a possible rotation period for this star. }
The values of inclination of the rotation axis for the seven stars with HARPS-N spectra are also listed in Table~\ref{Tab:param}.

\begin{figure}
\begin{center}
\hspace{-1cm}
\includegraphics[width=9.5cm]{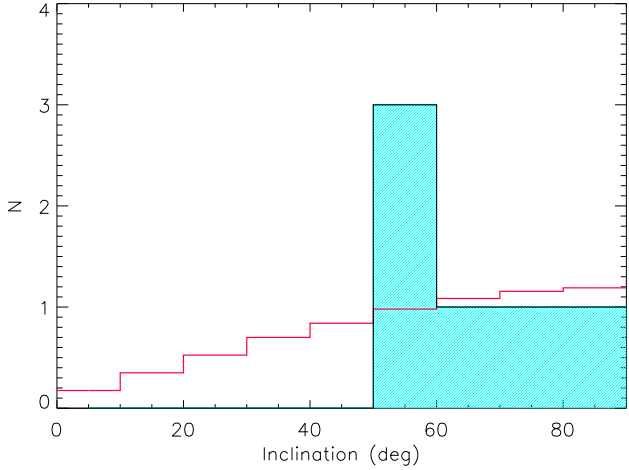}	
\vspace{0cm}
\caption{Distribution of the inclination of the rotation axis for the seven stars in Table\,\ref{Tab:param} (light-blue histogram). The red histogram shows the expected distribution for a three-dimensional
uniform orientation of the spin vectors. }
\label{fig:hist_sini}
\end{center}
\end{figure}

The statistical sample is too small to study the distribution of the inclinations of the stellar spins and investigate their degree of alignment, as pointed out, e.g., 
by \citet{Corsaro2017}, whose simulations show that a sample 2--3 times larger is required for obtaining a high significance of any eventual spin alignment. 
However, we wanted to compare the values of $i$ that we find for the seven stars in Table\ref{Tab:param} with those expected from a random 3D distribution of the 
rotation axes, in a similar way to what \citet{Corsaro2017} did for  giant stars in the old open clusters NGC\,6791 and NGC\,6819.
In their case, for both clusters the $i$ distribution was concentrated towards low values, i.e. it was very different from that expected from a random distribution 
of the rotation axes (see Figure 1 in their paper), suggesting a significant alignment of the stellar spins.
In our case (Fig.\,\ref{fig:hist_sini}), the values of inclination are all greater than 50 degrees and are compatible with a random distribution or, in any case, 
they do not allow us to draw firm conclusions. High- or medium-resolution spectra for a significant number of cluster members, as expected from future multi-object 
surveys like WEAVE \citep{WEAVE2022}, are needed to better investigate this aspect.
We note that \citet{Healy2021} find that isotropic spins or moderate alignment are both consistent with the $\sin i$ values they derived for members of Pleiades and 
Praesepe clusters. In a recent work \citep{Healy2023} found eight out of the ten investigated clusters having spin-axis orientations consistent with isotropy and two 
clusters whose distributions can be better described by an aligned fraction of stars combined with an isotropic distribution. However, they suggest this result can be 
influenced by systematic errors on $i$ and by the poor statistics.  

\subsection{Chromospheric activity}

\begin{figure}
\begin{center}
\includegraphics[width=8.5cm]{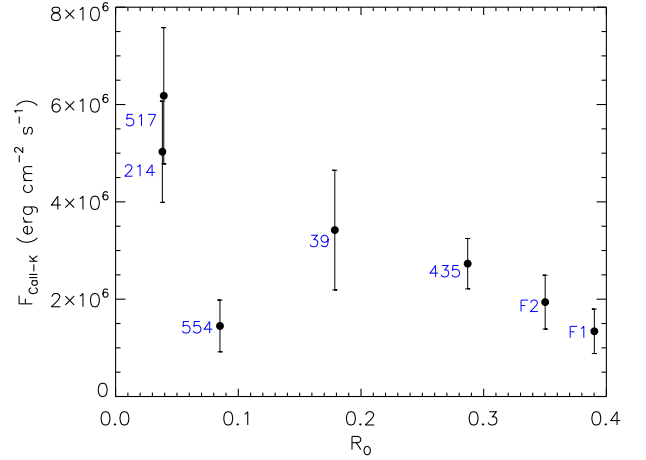}             
\vspace{0cm}
\caption{Line flux in the \ion{Ca}{\sc ii}-K line as a function of the Rossby number. }
\label{fig:FK_Prot}
\end{center}
\end{figure}

Correlations between diagnostics of chromospheric activity and \prot, or the Rossby number, have often been reported in the literature 
\citep[e.g.,][and references therein]{Douglas2014,Frasca2016,Han2023}. For rapidly rotating stars, a saturated regime, in which the activity level is constant 
with \prot, is clearly shown by the coronal X-ray emission \citep[e.g.][]{Pizzolato2003,Wright2011} but it is also seen at chromospheric level 
\citep[e.g.,][]{Douglas2014,Newton2017}. In particular, \citet{Douglas2014} found that H$\alpha$ activity is saturated for $R_{\rm O}<0.11$ for members of 
Hyades and Praesepe, while \citet{Newton2017} found saturation in their sample of nearby field M-type stars for $R_{\rm O}<0.2$.

Despite the small number of stars with measured H$\alpha$ and \ion{Ca}{\sc ii} fluxes, we wanted to see if there is any hint of a correlation with rotation 
in our data. As an example we show $F_{\rm CaII-K}$ as a function of $R_{\rm O}$ in Fig.~\ref{fig:FK_Prot}, which clearly shows the decline of flux with 
increasing $R_{\rm O}$. The only discrepant point is that of the hottest star, S\,554. Similar results are also obtained with \ion{Ca}{\sc ii}-H and H$\alpha$ 
as evidenced by the Pearson's correlation coefficients $\rho=-0.65$, $-0.71$, and $-0.74$, for the H$\alpha$, \ion{Ca}{\sc ii}-K, and H line, respectively.
Apart from the two F4V sources (S\,39 and S\,554), only the two ultrafast G-type stars (S\,214 and S\,517) should have a saturated chromospheric emission, 
on the basis of their Rossby number. These are the only stars in Table~\ref{Tab:param} for which a value of X-ray flux is reported in the literature. 
The values of $F_{\rm X}=3.36\times 10^{-13}$ and $3.65\times 10^{-13}$\,erg \,cm$^{-2}$s$^{-1}$ reported by \citet{Freund2022} translate, with the {\it Gaia} 
parallaxes and our values of \teff\ and $R_*$ (Table~\ref{Tab:param}),  into $\log(L_{\rm X}/L_{\rm bol})$ values of $-3.36$ and $-3.31$  for S\,214 and S\,517, 
respectively. These values are close to the saturation level of $-3.13$ reported by \citet{Wright2011}. 

\begin{figure*}
\begin{center}
\includegraphics[width=8.5cm]{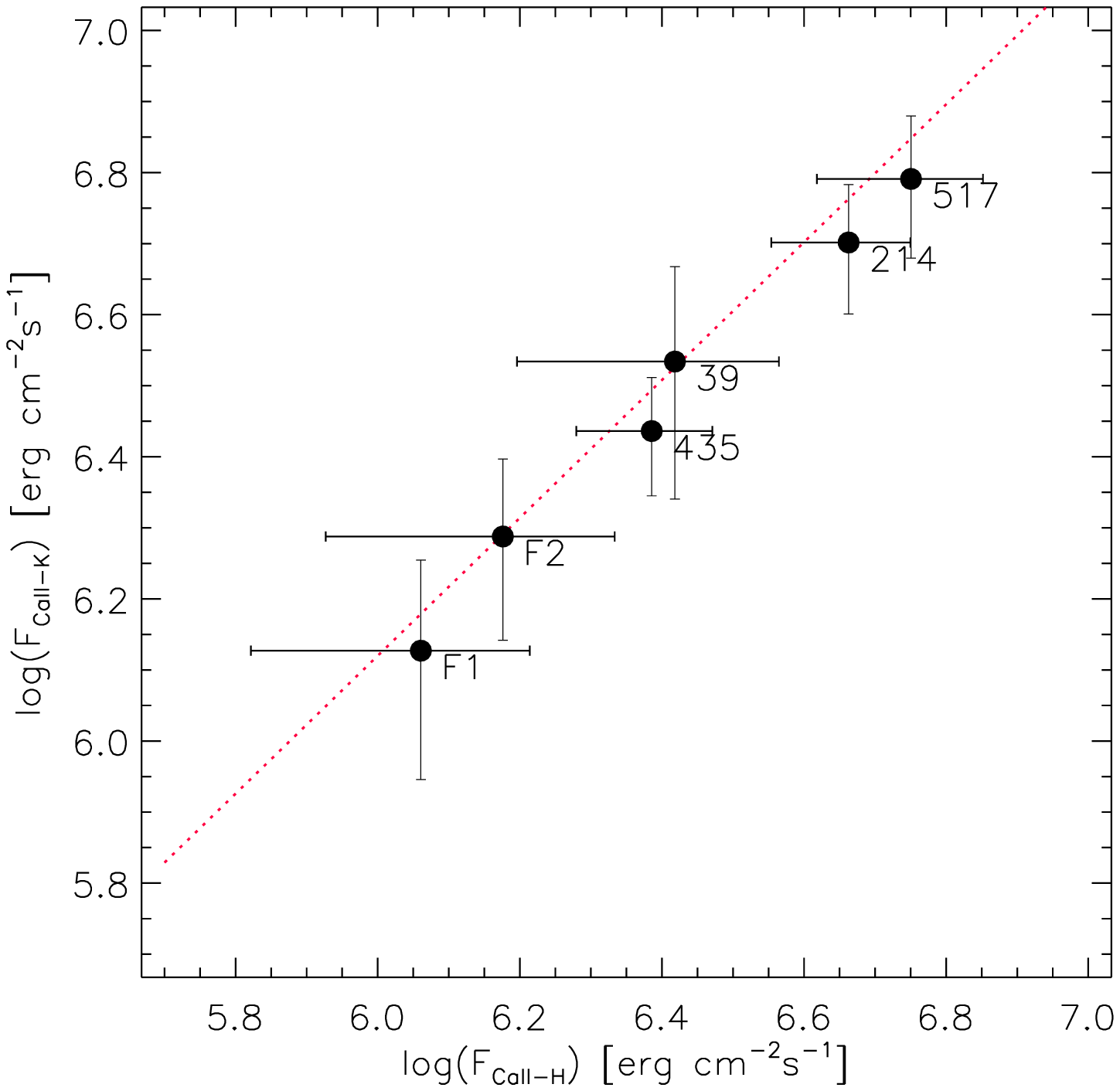}     
\includegraphics[width=8.5cm]{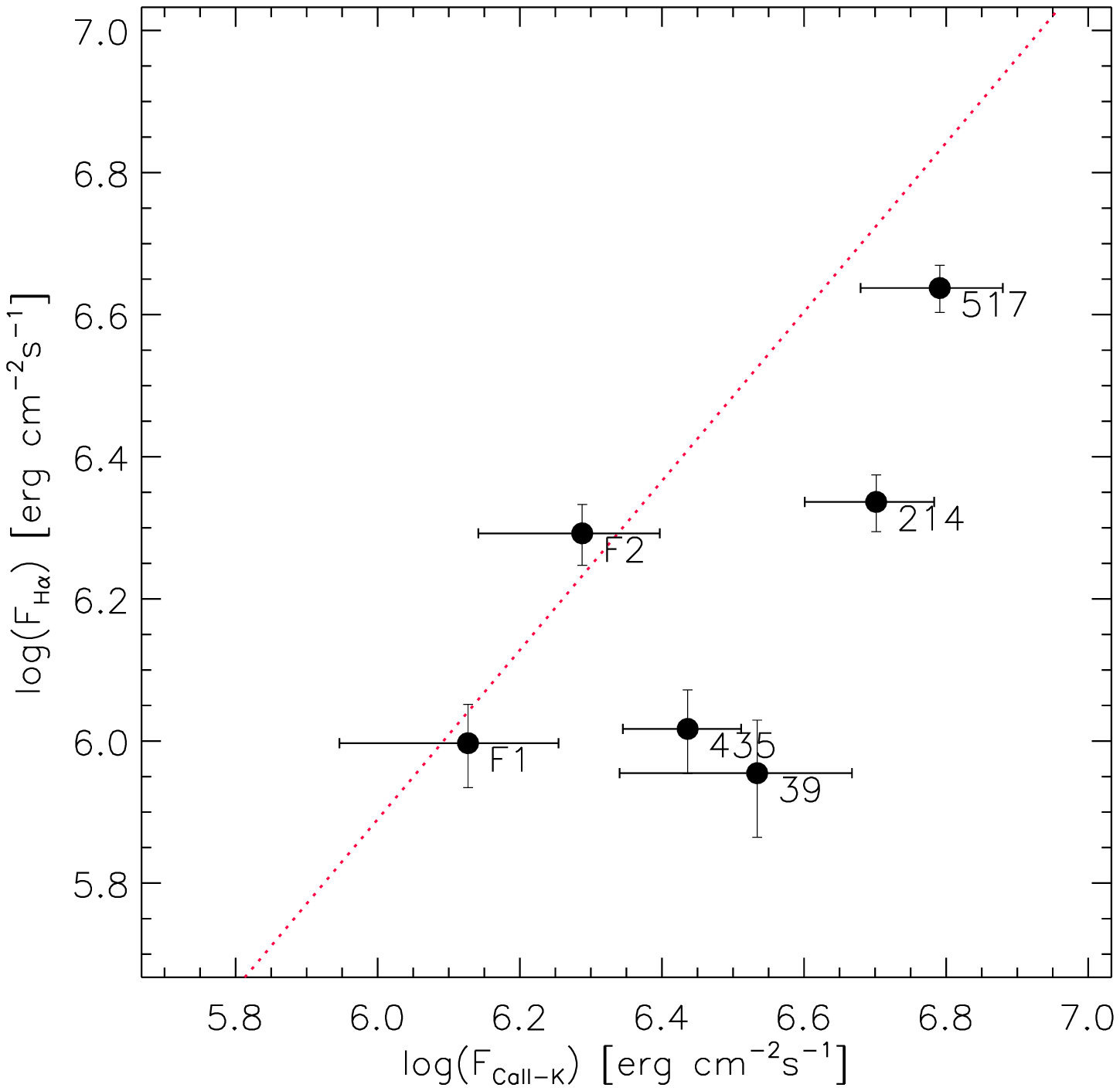}        
\vspace{0cm}
\caption{Flux--flux relationships between H\,\&\,K calcium lines ({\it left panel}) and H$\alpha$ ({\it right panel}). The red dotted lines represent the power laws 
that are best fitting the data of \citet{Martinez2011}. The blue dashed-line in the right panel represents the fit to the lower branch of the \citet{Martinez2011} 
H$\alpha$ data. }
\label{fig:flux-flux}
\end{center}
\end{figure*}

To further investigate the behaviour of the magnetic activity at chromospheric level we have compared the different diagnostics by means of flux--flux diagrams.
These diagrams are shown, in a logarithm scale, in Fig.~\ref{fig:flux-flux} along with the power-law relationships found by \citet{Martinez2011} from their large 
sample of active FGKM stars.
We note that all the members lie, within the error bars, over the flux--flux relation for the \ion{Ca}{ii}-K versus \ion{Ca}{ii}-H, while the H$\alpha$ fluxes, 
which are displayed versus \ion{Ca}{ii}-K ones in the right panel of Fig.~\ref{fig:flux-flux}, are more scattered.
This is consistent with the results of \citet{Martinez2011}, who found a larger scatter in the H$\alpha$--\ion{Ca}{ii} flux diagrams compared to diagrams based 
on different \ion{Ca}{ii} lines. Furthermore, they  distinguished two different regimes in the H$\alpha$--\ion{Ca}{ii} flux diagrams, with the very active late-K 
and M-type stars significantly deviating from the main behaviour and lying in an upper branch. They argued that these are stars with a saturated X-ray flux.
The stars in our sample are all earlier than those in the upper branch of the flux--flux diagrams of \citet{Martinez2011} and only the ultrafast G-type stars 
S\,214 and S\,527 display an activity level close to saturation. The remaining GK-type stars should not be saturated, on the basis of their Rossby number and the 
above mentioned correlations.
However, we note that the two K-type stars F1 and F2 lie right on the average power-law relation (red dotted line in Fig.~\ref{fig:flux-flux}), while the remaining 
stars are placed systematically lower, but are closer to the lower branch fit (blue dashed line in Fig.~\ref{fig:flux-flux}). The latter was calculated by 
\citet{Martinez2011} by excluding the stars with saturated X-ray and H$\alpha$ emission from the fit.

\section{Summary and conclusions}\label{Sec:conclusions}

In this work we have revisited ASCC\,123. It is a little-studied young cluster, with no detailed spectroscopic investigation apart our
recent study \citep{Frasca2019}.
In that paper we determined the main properties of the cluster such as distance, age and chemical composition by combining $Gaia$ data with high-resolution 
spectroscopy taken with HARPS-N at the TNG telescope. In the present work we have focused on rotation and magnetic activity. With this aim we retrieved {\it TESS} 
photometry for all the 55 cluster members reported in \citet{Cantat2018}. Among them, seven FGK-type stars were observed spectroscopically in our previous work. 
We performed the analysis of the light curves of the cluster members obtaining the rotational periods and amplitudes for 29 objects. By comparing the distributions 
of period and amplitude versus colour index with those of some well-studied clusters, we infer a gyrochronogical age for ASCC\,123 similar to that of the Pleiades, 
which is consistent with our precedent estimates based on the isochrone fitting and abundance of lithium. No clear correlation between the variation amplitude and 
\prot\ is observed, while a Spearman's coefficient $\rho=-0.16$ between amplitude and Rossby number suggests only a marginal correlation of these variables.
Additionally, for the seven stars with spectra we have derived the inclination of the rotational axis. Although our sample is not statistically significant, the 
distribution of inclinations seems to be random, which would be in agreement with that observed in other young open clusters. Finally, we studied the level of magnetic 
activity of these stars from the H$\alpha$ and \ion{Ca}{II} H\&K lines. We find that the chromospheric activity indicators display a better correlation 
with the Rossby number ($\rho$ in the range [$-0.65,-0.74$]). Moreover, they follow the general trends observed in other active FGKM stars in the flux--flux diagrams.

\section*{Acknowledgments}
We thank the anonymous referee for a careful reading of the manuscript and valuable comments and suggestions.
This research used the facilities of the Italian Center for Astronomical Archive (IA2) operated by INAF at the Astronomical Observatory of Trieste.
This research has made use of the SIMBAD database and VizieR catalogue access tool,
operated at CDS, Strasbourg, France.\\
This work has made use of data from the European Space Agency (ESA)
mission \gaia\ ({\tt https://www.cosmos.esa.int/gaia}), processed by
the {\it Gaia} Data Processing and Analysis Consortium (DPAC,
{\tt https://www.cosmos.esa.int/web/gaia/dpac/consortium}). Funding
for the DPAC has been provided by national institutions, in particular
the institutions participating in the {\it Gaia} Multilateral Agreement.\\
This paper includes data collected by the {\it TESS} mission which are publicly available from the Mikulski Archive for Space Telescopes (MAST). 
Funding for the {\it TESS} mission is provided by the NASA's Science Mission Directorate.
This work has been supported by the PRIN-INAF 2019 STRADE (Spectroscopically TRAcing the Disk dispersal Evolution) and by the Large Grant INAF YODA 
(YSOs Outflow, Disks and Accretion),

\section*{Data Availability}
The spectroscopic and ground-based photometric data underlying this paper will be shared on reasonable request to the corresponding author. 
{\it TESS} photometric data are available at {\tt https://archive.st sci.edu/}. 

\bibliographystyle{mnras}
\bibliography{asc123_mnras.bib}

\label{lastpage} 

\newpage

\begin{appendix}

\section{Ground-based photometry}
\label{Sec:Ground_Phot}

\begin{figure}
\begin{center}
\hspace{-0.7cm}
\includegraphics[width=5.6cm]{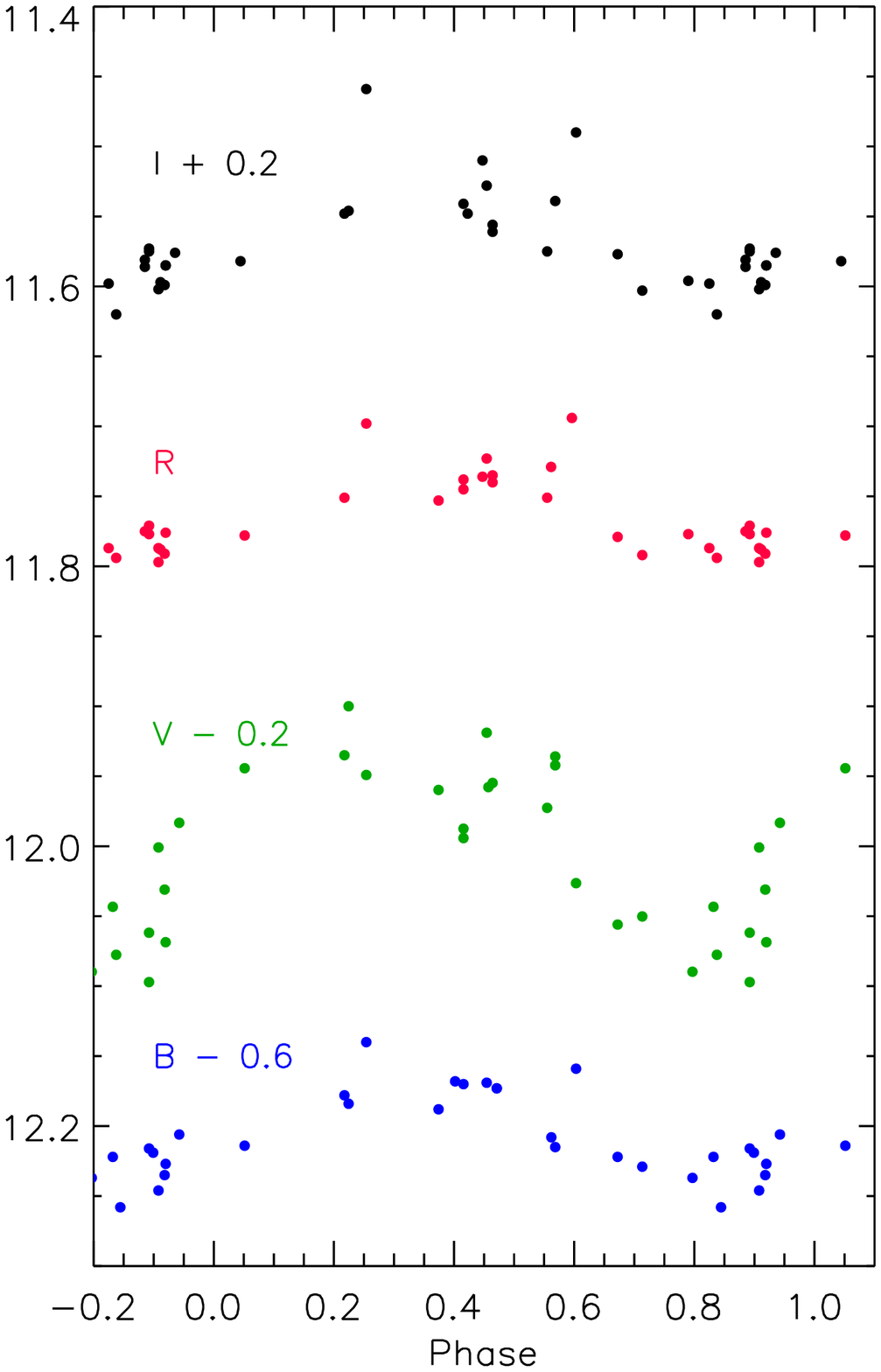}	
\hspace{-0.3cm}
\includegraphics[width=3.45cm]{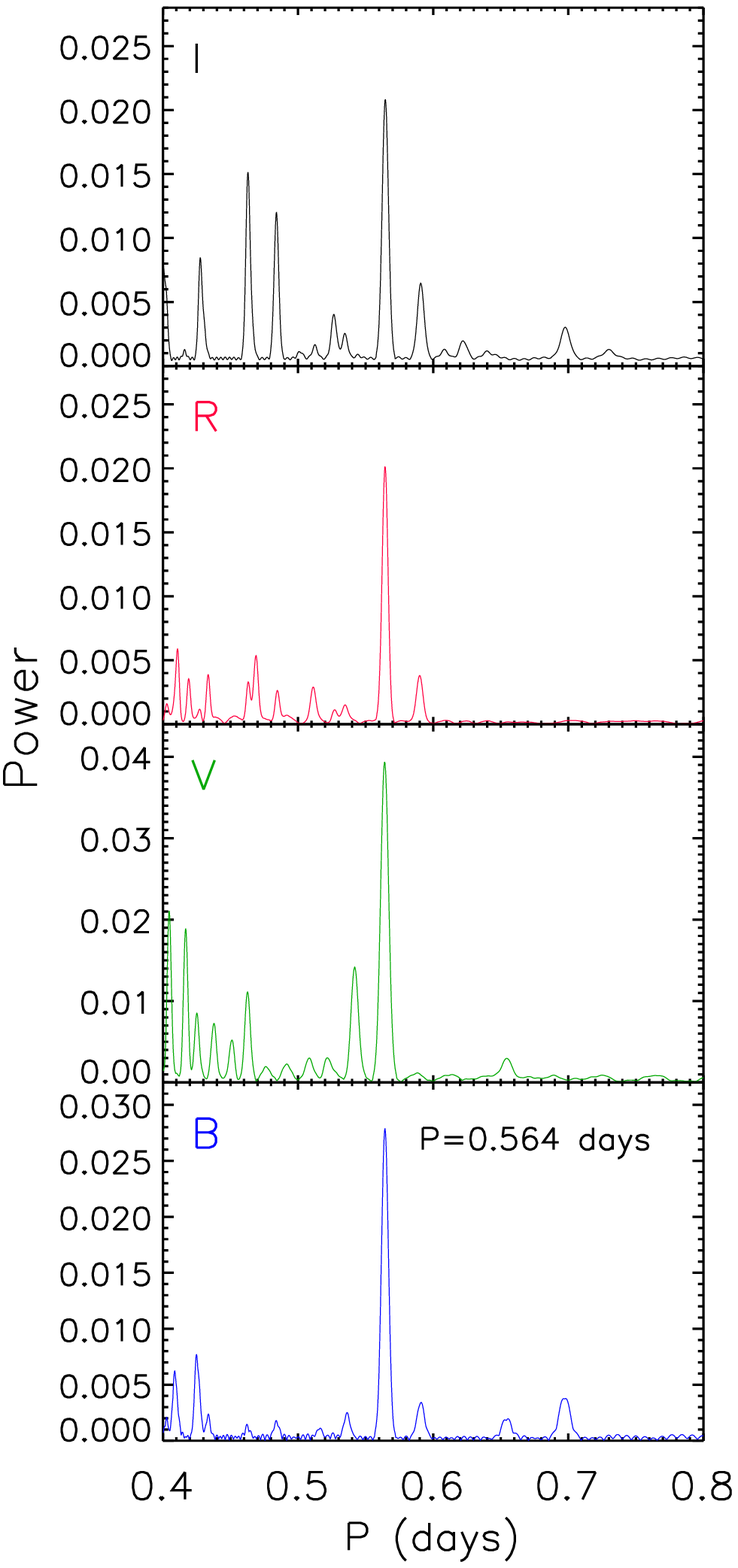}	
\vspace{0cm}
\caption{{\it (Left panel)} Phased $BVR_{\rm C}I_{\rm C}$ light curves of S\,214 collected at OACT. {\it (Right panels)} Cleaned periodograms.  }
\label{fig:SLN_LC}
\end{center}
\end{figure}

As mentioned in Sect.\,\ref{Subsec:Obs_photo}, we collected multiband $BVR_{\rm C}I_{\rm C}$ photometry of the three G-type stars S\,214, S\,435, and S\,517 at OACT. 
Despite the lower photometric precision, compared with {\it TESS} data, and the irregular sampling of our photometry, we were able to detect the rotational 
modulation and derive the period for the two ultrafast rotators (S\,214 and S\,517). The light curves in the $BVR_{\rm C}I_{\rm C}$ bands of S\,214 are shown in 
Fig.\,\ref{fig:SLN_LC}. The peak at a rotation period of 0.564 days is the highest one in the cleaned power spectrum of each band and is also recovered from the 
{\it TESS} photometry of this source (see Sect.~\ref{Sec:TESS}). The amplitudes of the $BVR_{\rm C}I_{\rm C}$ light curves are about 0.11, 0.15, 0.08, 
and 0.07\,mag, respectively.
The average magnitudes of these stars in the Johnson-Cousins bands from our OACT photometry are listed in Table\,\ref{Tab:Stand_phot}.

\setlength{\tabcolsep}{3pt}

\begin{table}
\caption{Mean Johnson-Cousins magnitudes of the three G-type members observed at OACT.}
\begin{center}
\begin{tabular}{lcccccccc}
\hline
\noalign{\smallskip}
ID      & $B$ & err& $V$  & err & $R_{\rm C}$ & err & $I_{\rm C}$ & err \\
        & \multicolumn{2}{c}{(mag)} & \multicolumn{2}{c}{(mag)} & \multicolumn{2}{c}{(mag)} & \multicolumn{2}{c}{(mag)}  \\  
\noalign{\smallskip}
\hline
\noalign{\smallskip}
214  & 12.866 & 0.031 & 12.192 & 0.051 & 11.769 & 0.026 & 11.403 & 0.022 \\ 
435  & 12.993 & 0.033 & 12.242 & 0.035 & 11.922 & 0.033 & 11.527 & 0.029 \\ 
517  & 13.093 & 0.028 & 12.199 & 0.036 & 11.773 & 0.025 & 11.417 & 0.022 \\
\noalign{\smallskip}
\hline
\noalign{\smallskip}
\end{tabular}
\label{Tab:Stand_phot}
\end{center}
\end{table}

\section{Additional figures}

\begin{figure}
\hspace{-0.5cm}
\includegraphics[width=9.4cm]{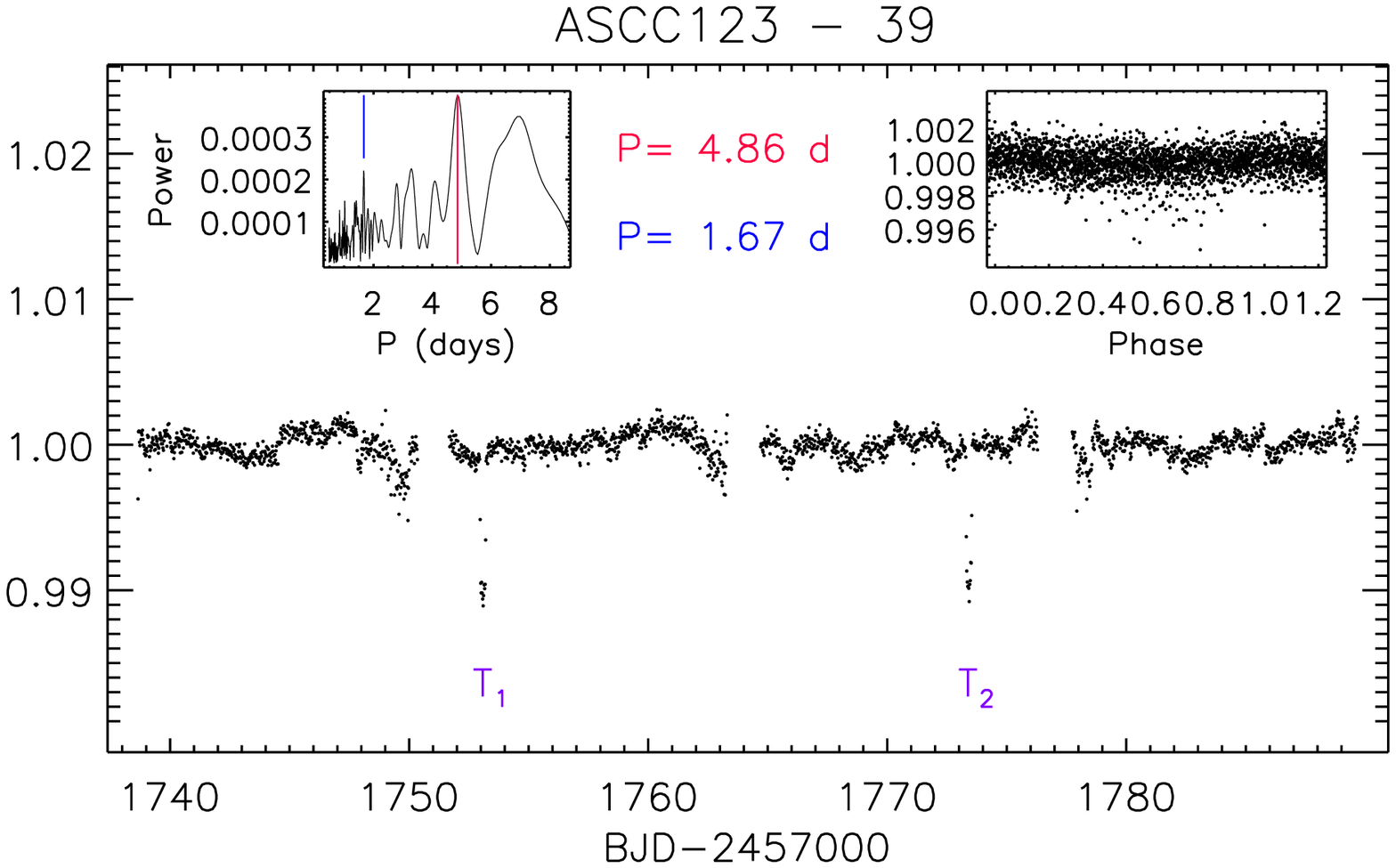}	
\vspace{-0.5cm}
\caption{{\it TESS} light curve of S\,39 (TIC\,64837857) in 2019 (sectors 16 and 17). The two transits are marked with T$_1$ and T$_2$.  
The inset in the upper left corner shows the cleaned periodogram of the data excluding the transits; the period corresponding to the
maximum power is marked with a vertical red line and is written next to the box. The blue vertical line marks the possible 
rotational period of about 1.67 days. The inset in the upper right corner displays the data phased with this period.}
\label{fig:TESS_ASC39}
\end{figure}

\begin{figure}
\begin{center}
\includegraphics[width=8.cm]{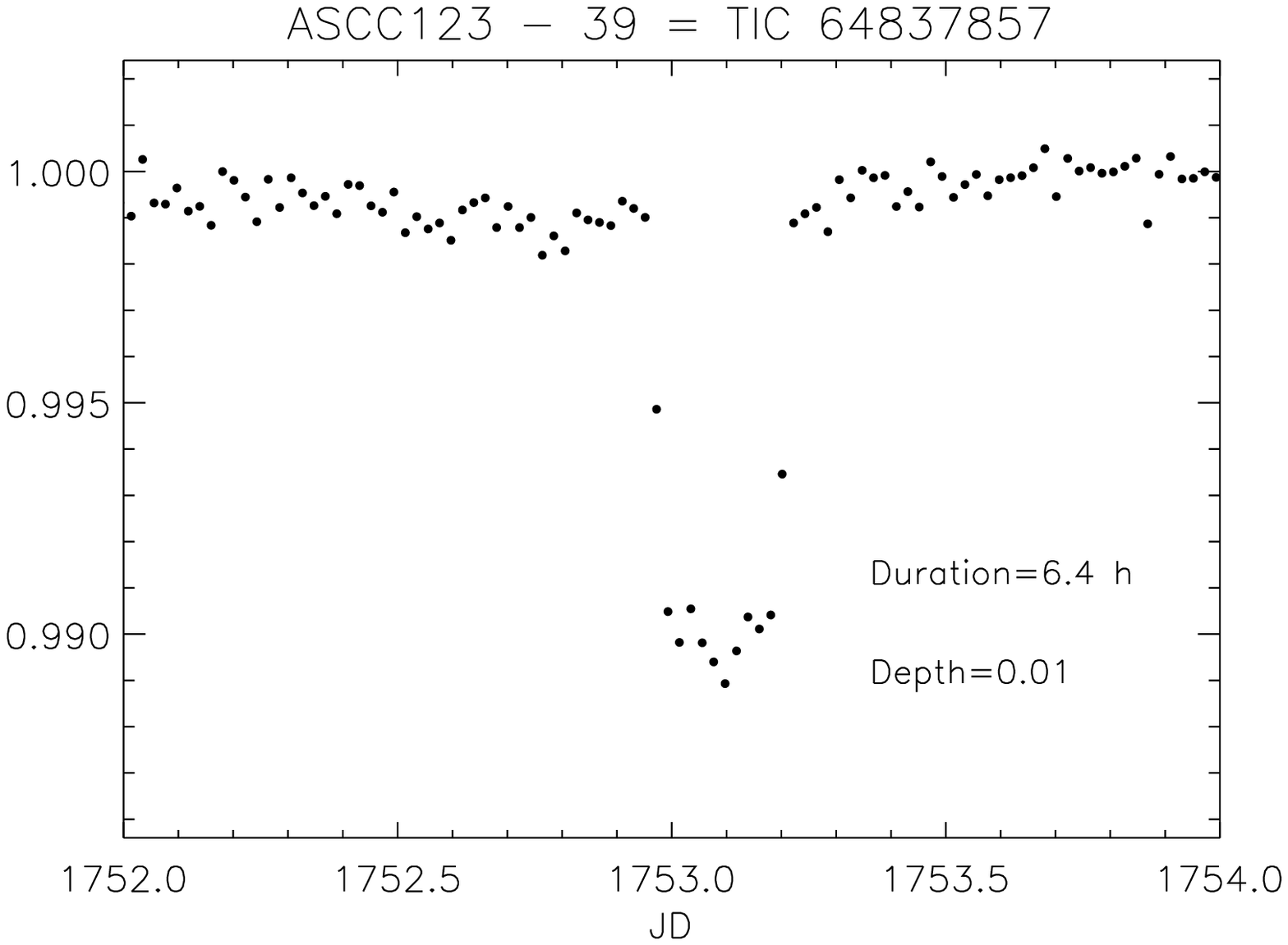}	
\hspace{-0.5cm}
\includegraphics[width=8.cm]{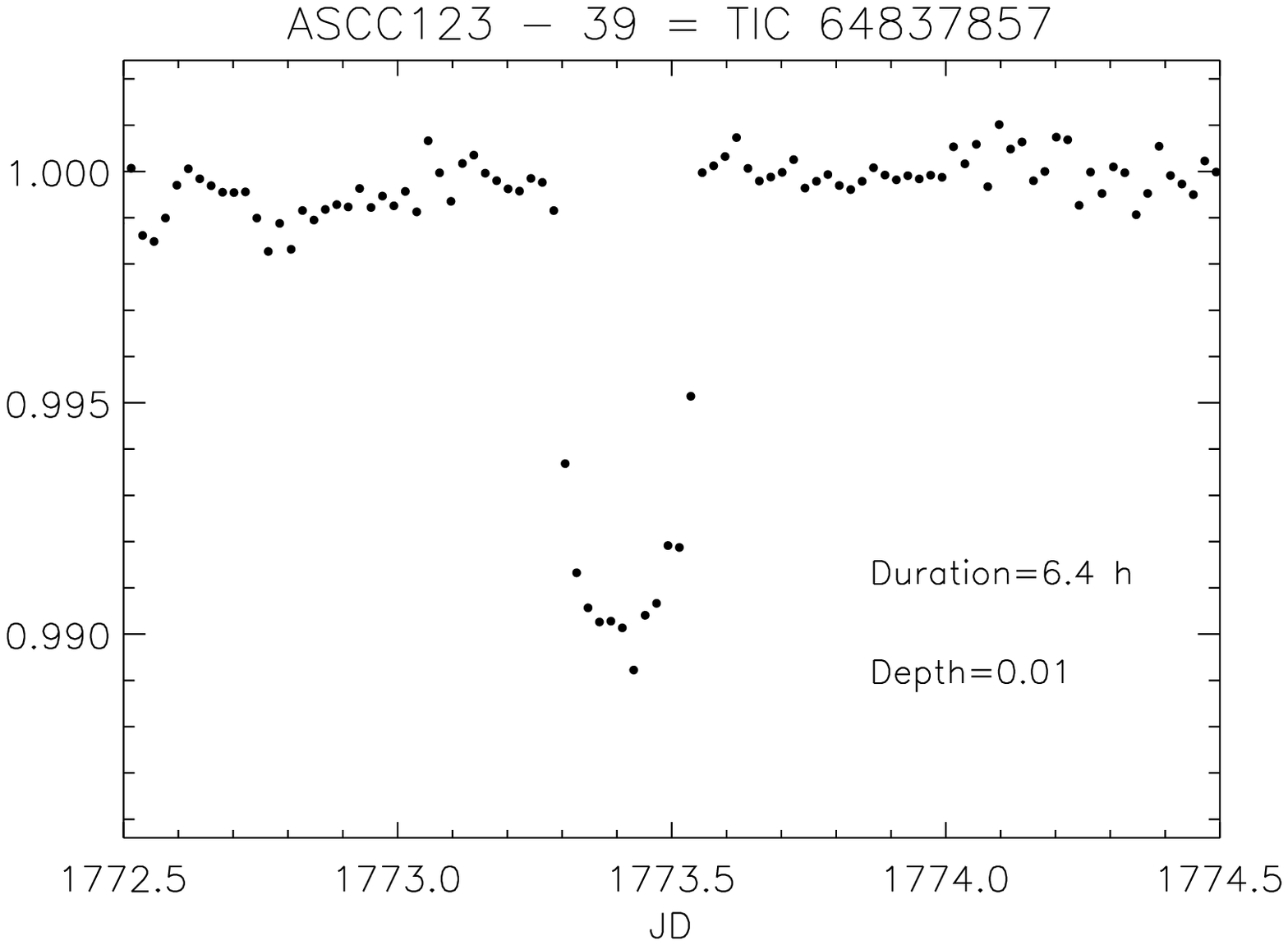}	
\vspace{0cm}
\caption{Zoom of the two transits detected in the {\it TESS} light curve of S\,39 (TIC\,64837857) in sectors 16 ({\it upper panel}) and 
17 ({\it lower panel}). }
\label{fig:TESS_Transits}
\end{center}
\end{figure}

\begin{figure}
\hspace{-0.5cm}
\includegraphics[width=9.4cm]{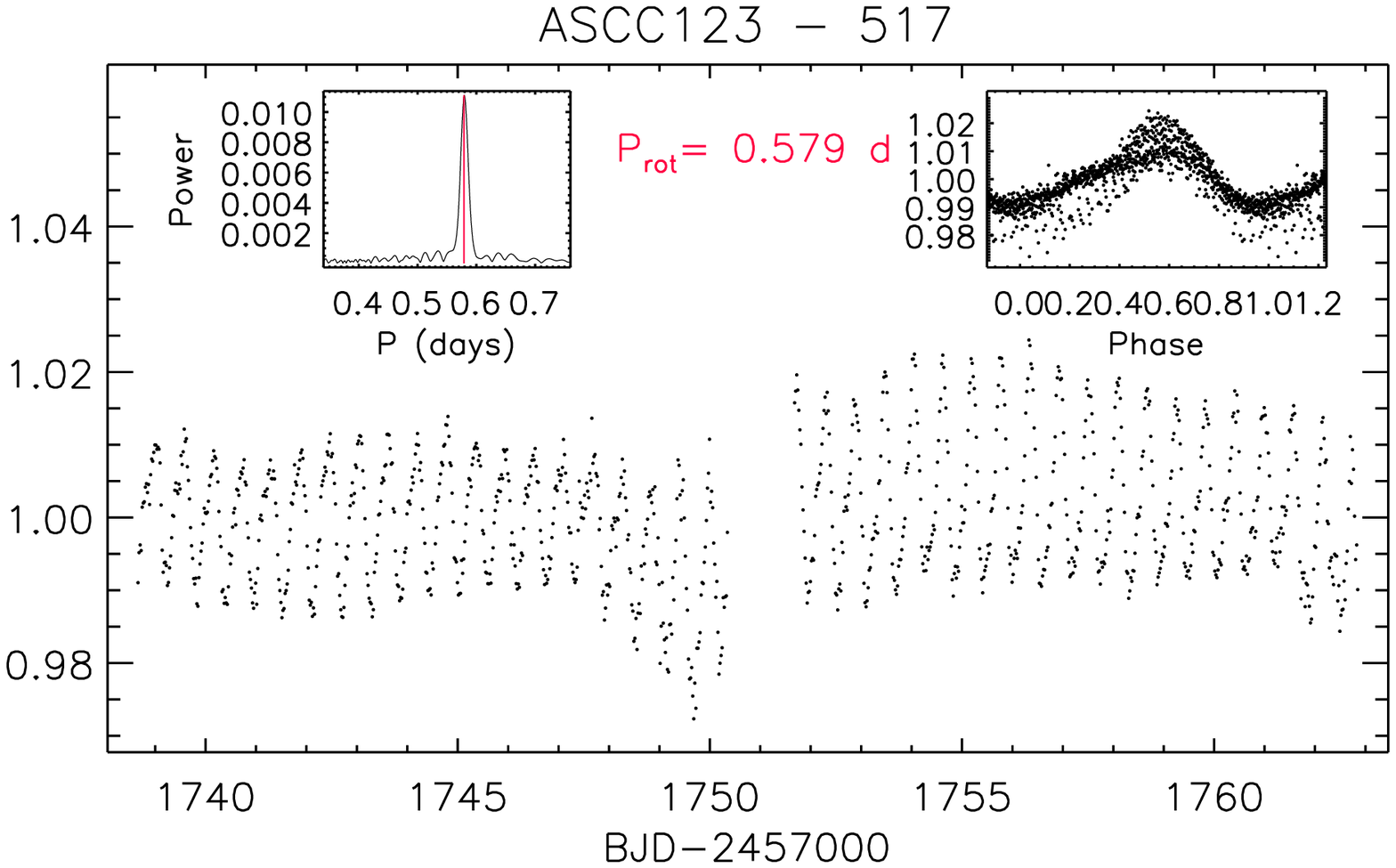}	
\vspace{-0.5cm}
\caption{{\it TESS} light curve of S\,517 (TIC\,428274538) in 2019 (Sector 16). The inset in the upper left corner shows the 
cleaned periodogram of these data; the rotation period is marked with a vertical red line. The inset in the upper right corner
displays the data phased with this period.}
\label{fig:TESS_ASC517}
\end{figure}

\begin{figure}
\hspace{-0.5cm}
\includegraphics[width=9.4cm]{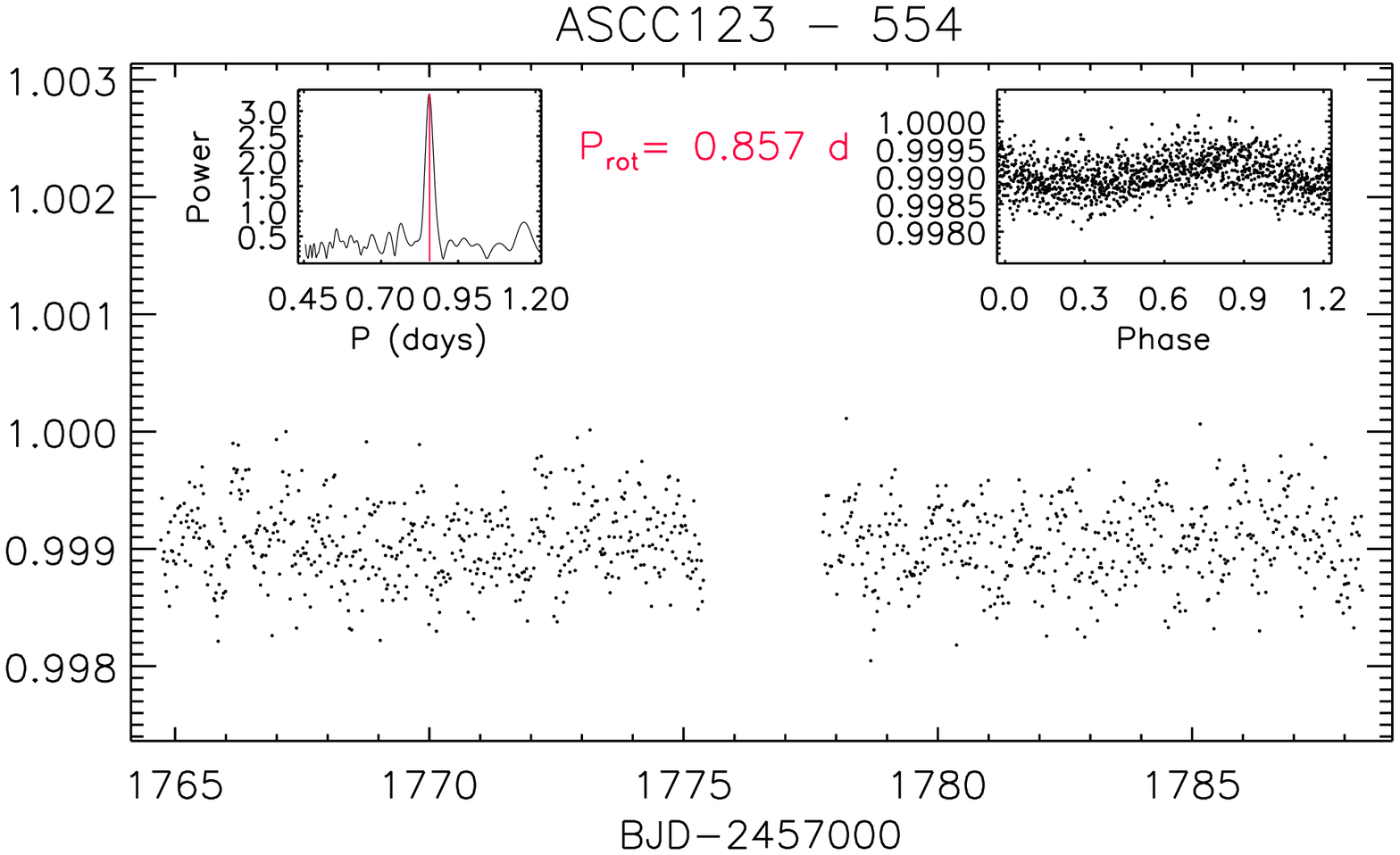}	
\vspace{-0.5cm}
\caption{{\it TESS} light curve of S\,554 (TIC\,361944360) in 2019 (sector 17). The figure layout is the same as Fig.~\ref{fig:TESS_ASC517}.
}
\label{fig:TESS_ASC554}
\end{figure}

\begin{figure}
\hspace{-0.5cm}
\includegraphics[width=9.4cm]{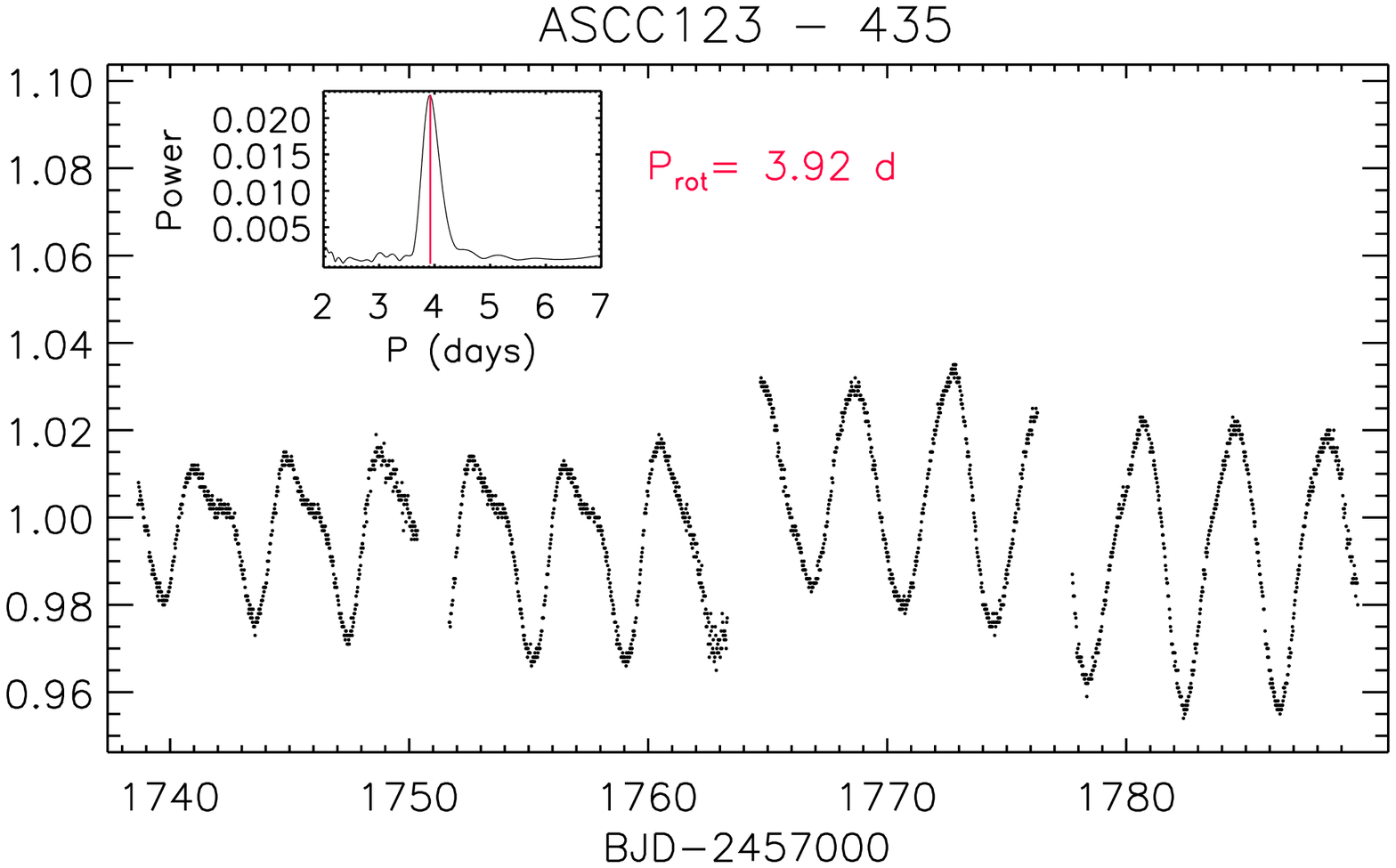}	
\vspace{-0.5cm}
\caption{{\it TESS} light curve of S\,435 (TIC\,388696341) in 2019 (sectors 16 and 17). The inset in the upper left corner shows the 
cleaned periodogram of these data; the rotation period is marked with a vertical red line. }
\label{fig:TESS_ASC435}
\end{figure}

\begin{figure}
\hspace{-0.5cm}
\includegraphics[width=9.4cm]{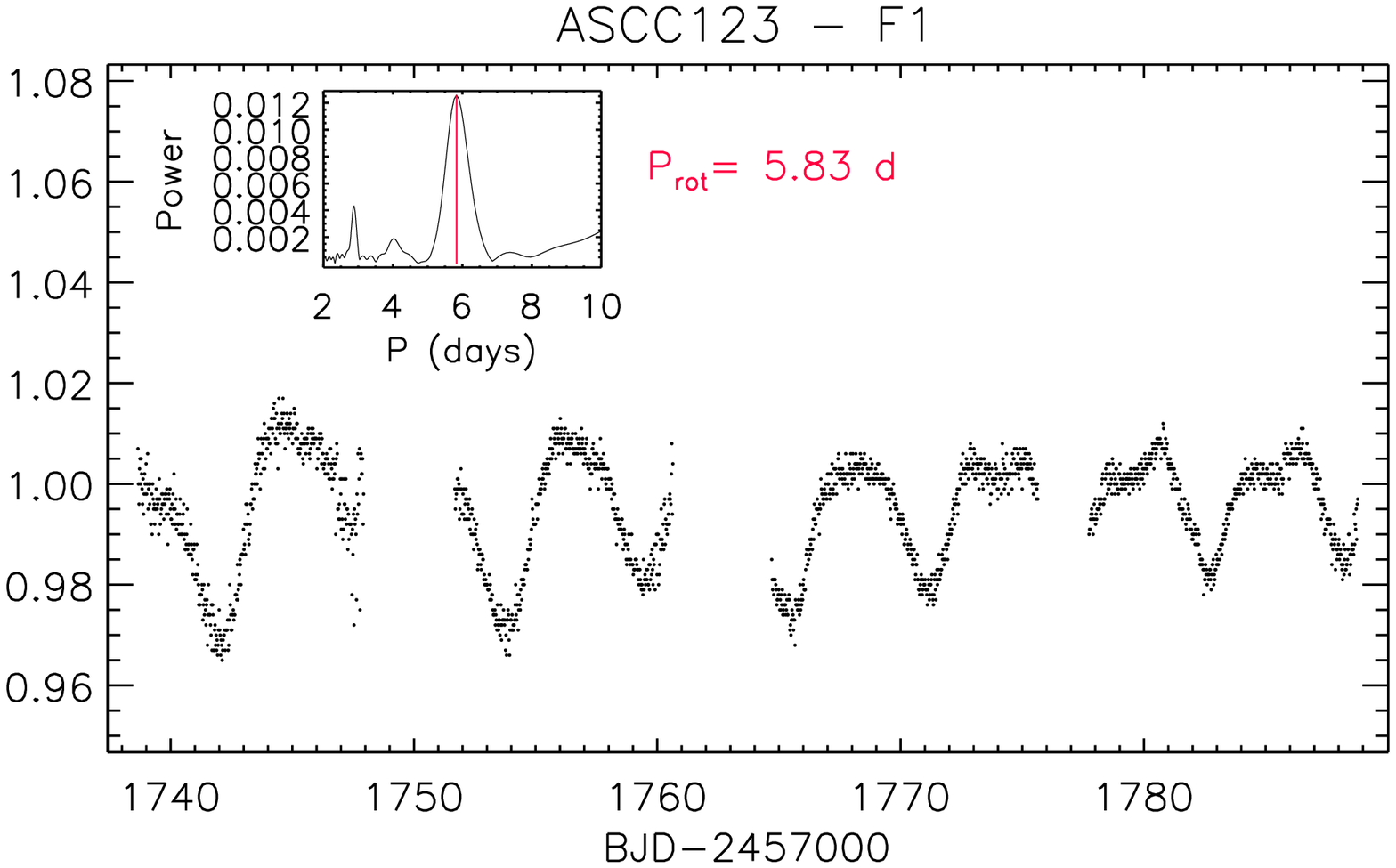}	
\vspace{-0.5cm}
\caption{{\it TESS} light curve of F1 (TIC\,66539637) in 2019 (sectors 16 and 17). The figure layout is the same as Fig.~\ref{fig:TESS_ASC435}.
}
\label{fig:TESS_F1}
\end{figure}

\begin{figure}
\hspace{-0.5cm}
\includegraphics[width=9.4cm]{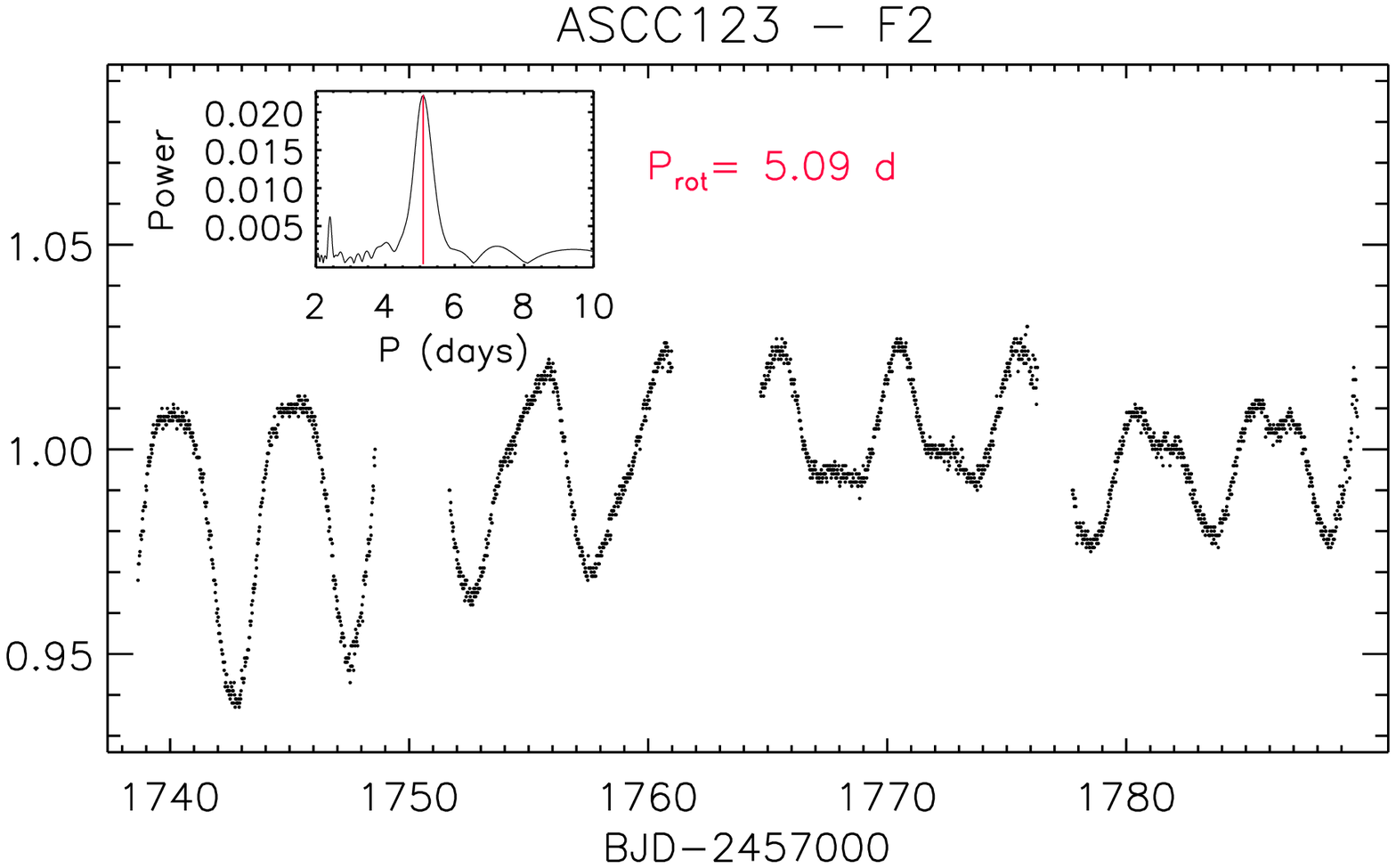}	
\vspace{-0.5cm}
\caption{{\it TESS} light curve of F2 (TIC\,64077901) in 2019 (sectors 16 and 17). The figure layout is the same as Fig.~\ref{fig:TESS_ASC435}.
}
\label{fig:TESS_F2}
\end{figure}

\begin{figure*}
\begin{center}
\includegraphics[width=5.9cm]{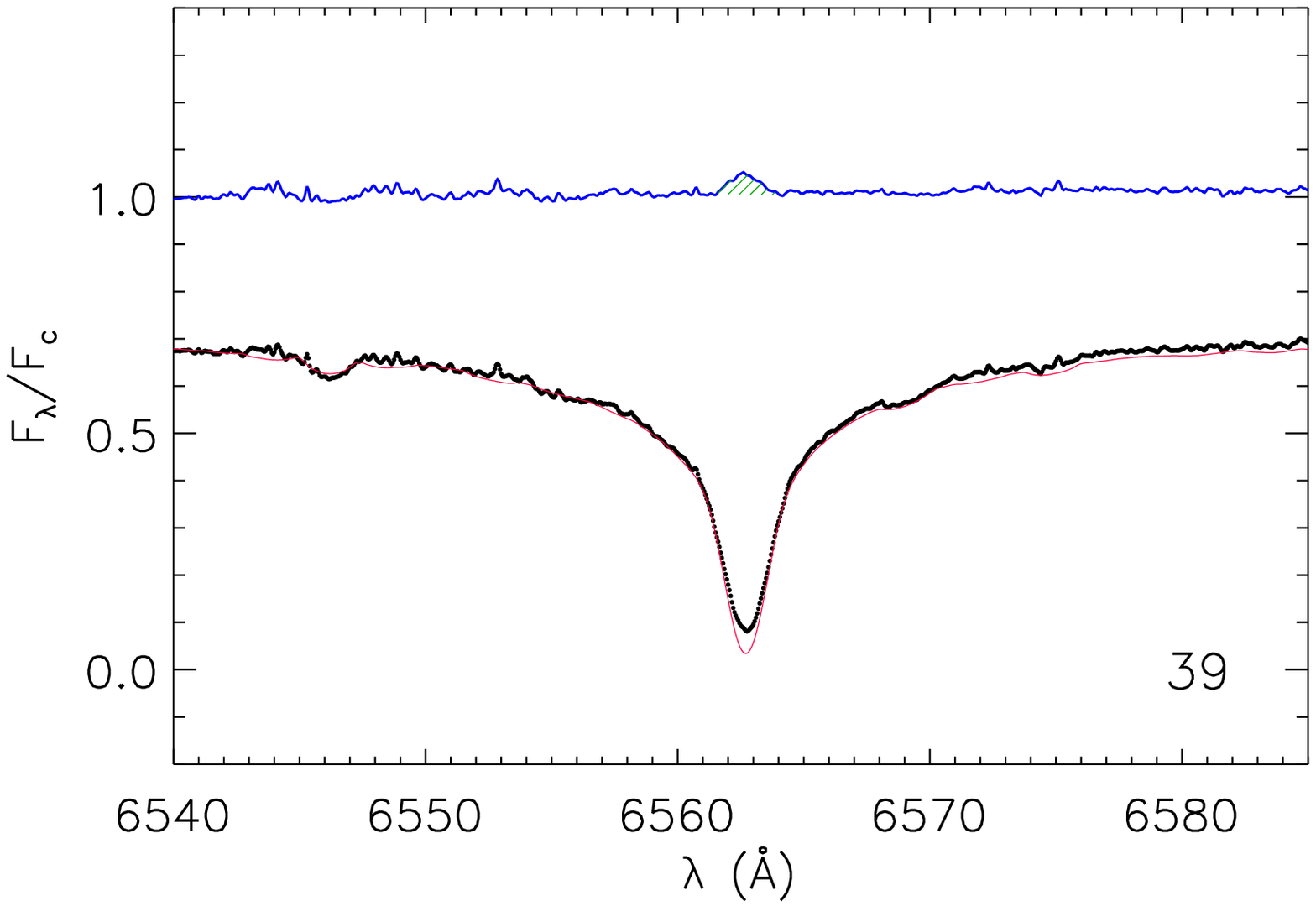}	
\hspace{-.5cm}
\includegraphics[width=5.9cm]{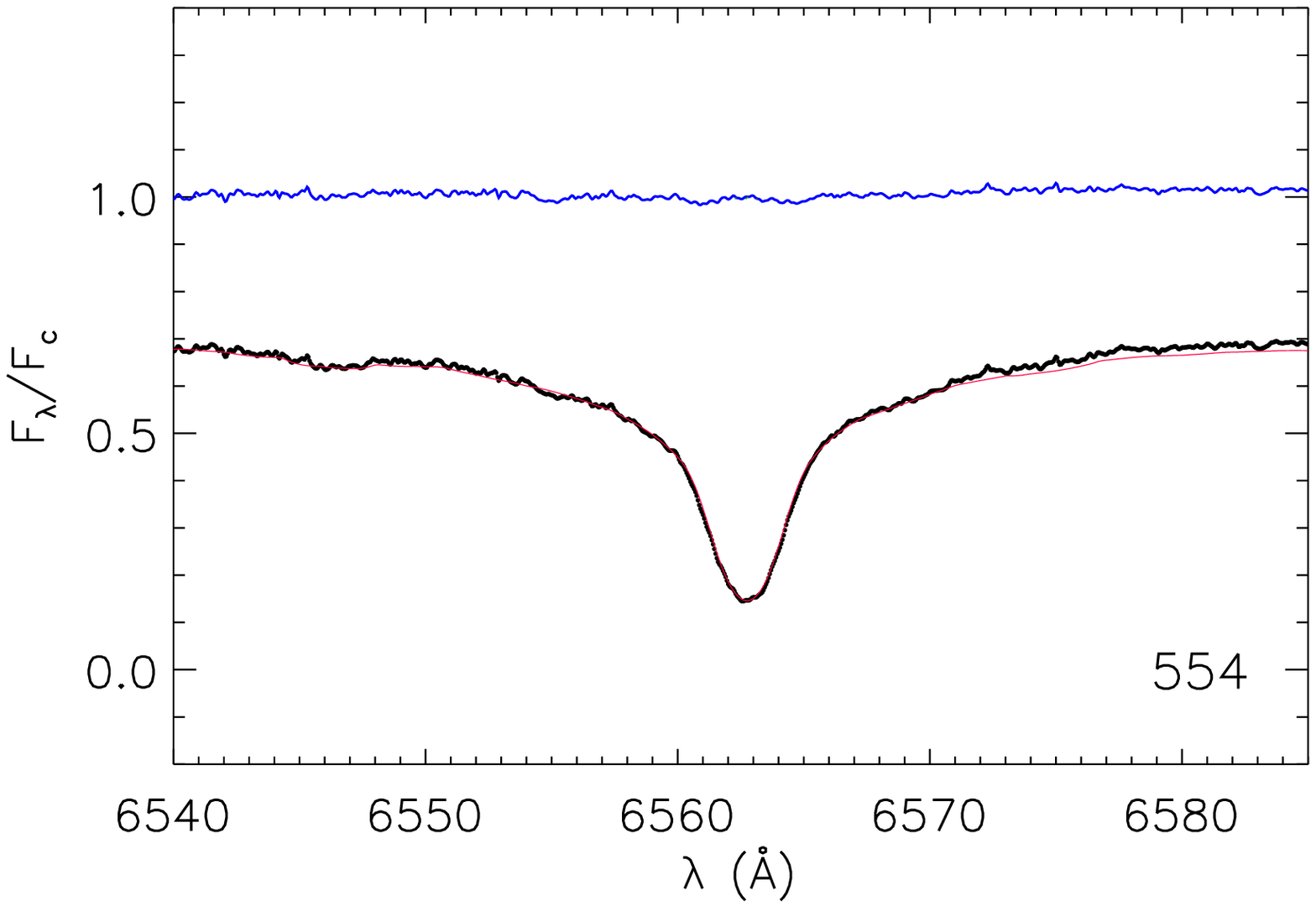}        
\hspace{-.5cm}
\includegraphics[width=5.9cm]{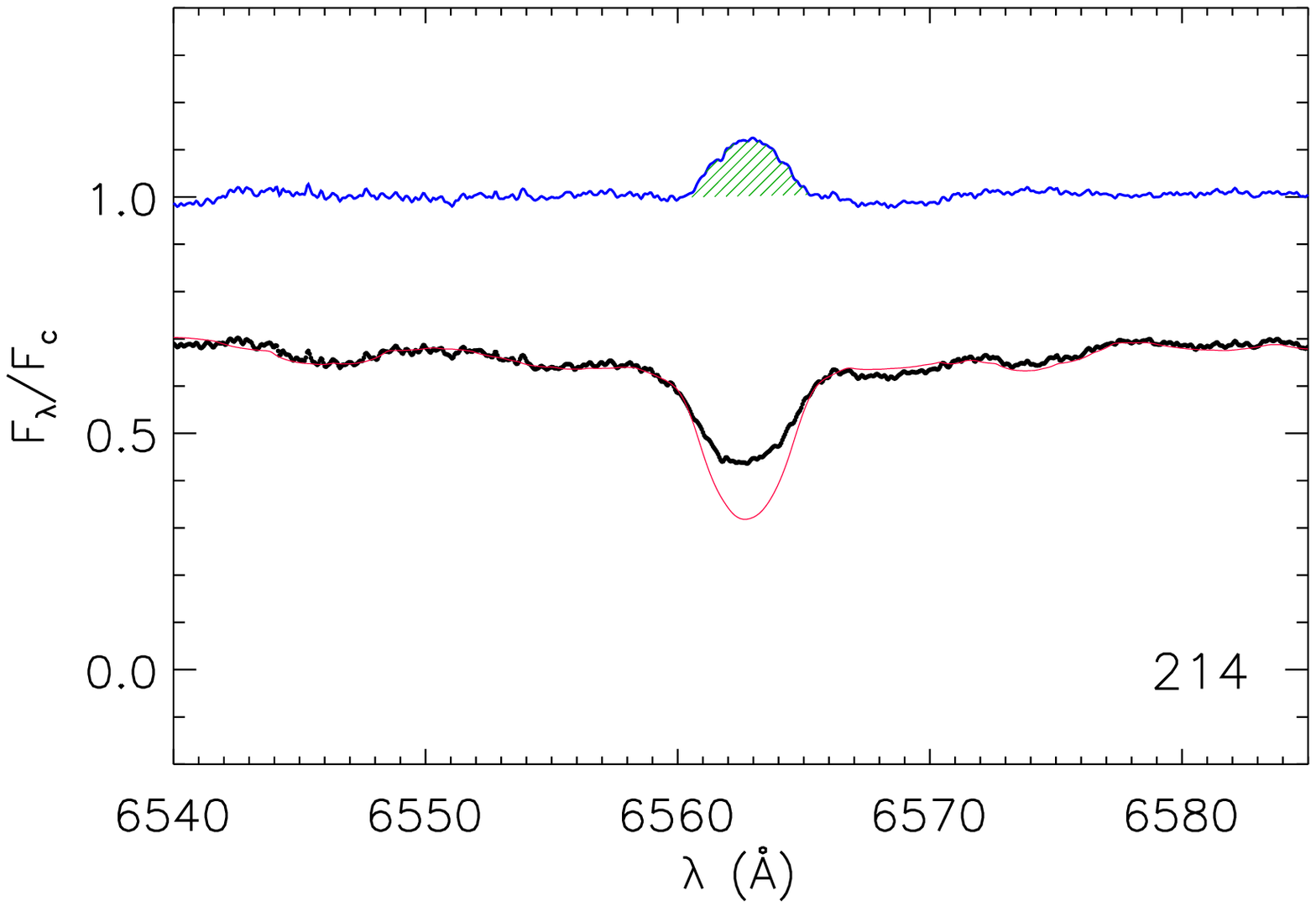}        
\hspace{-.5cm}
\includegraphics[width=5.9cm]{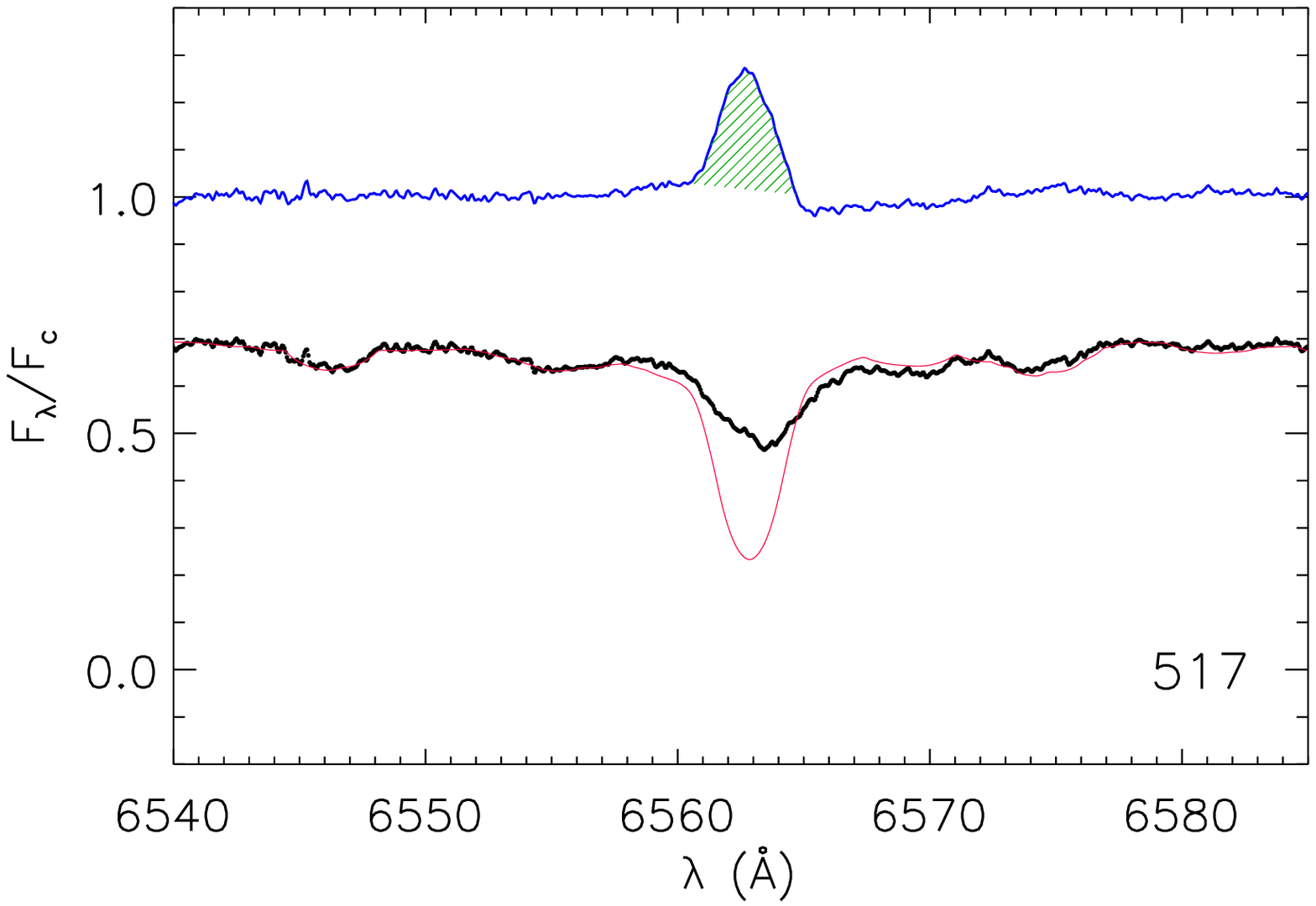}        
\hspace{-.5cm}
\includegraphics[width=5.9cm]{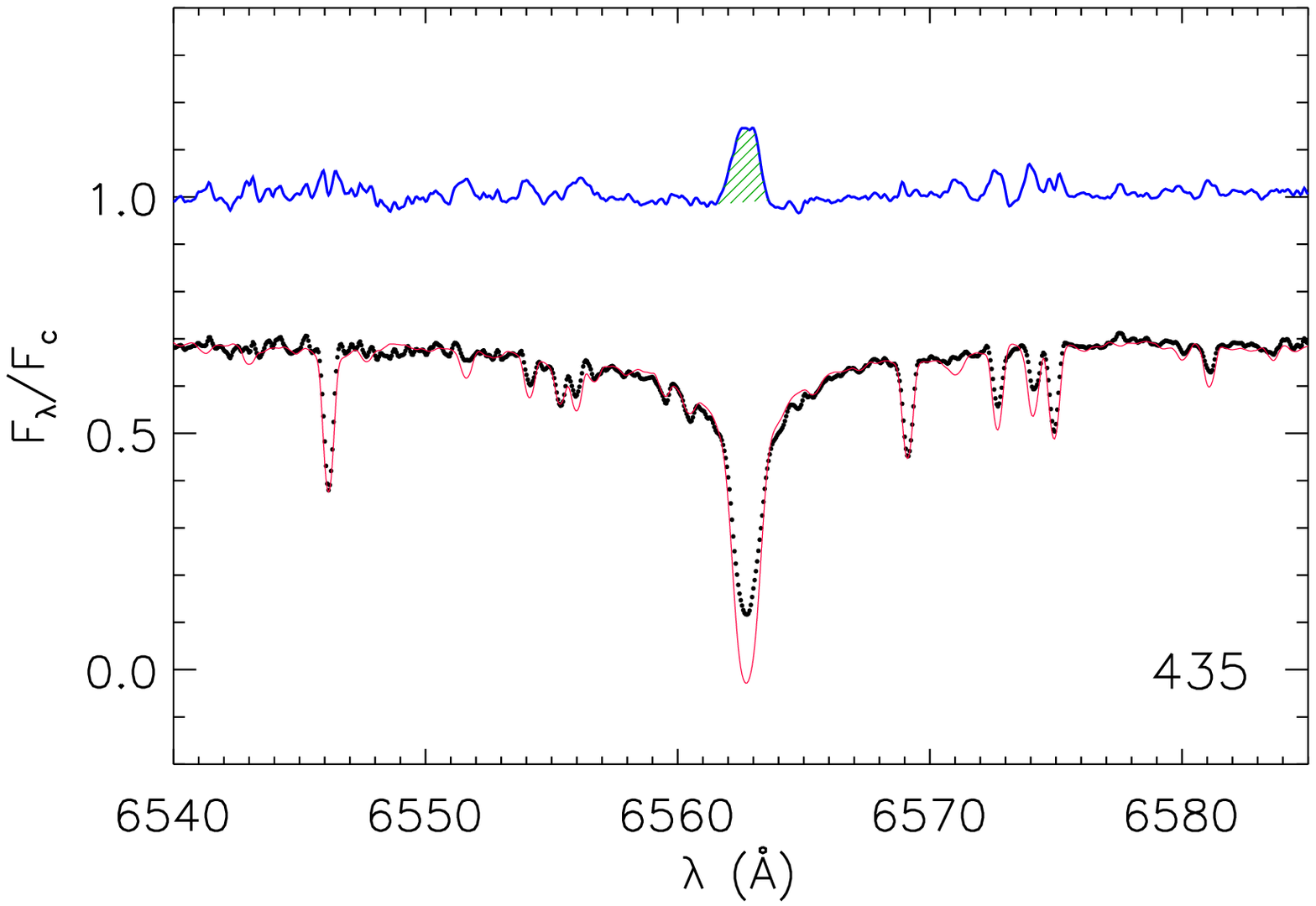}        
\hspace{-.5cm}
\includegraphics[width=5.9cm]{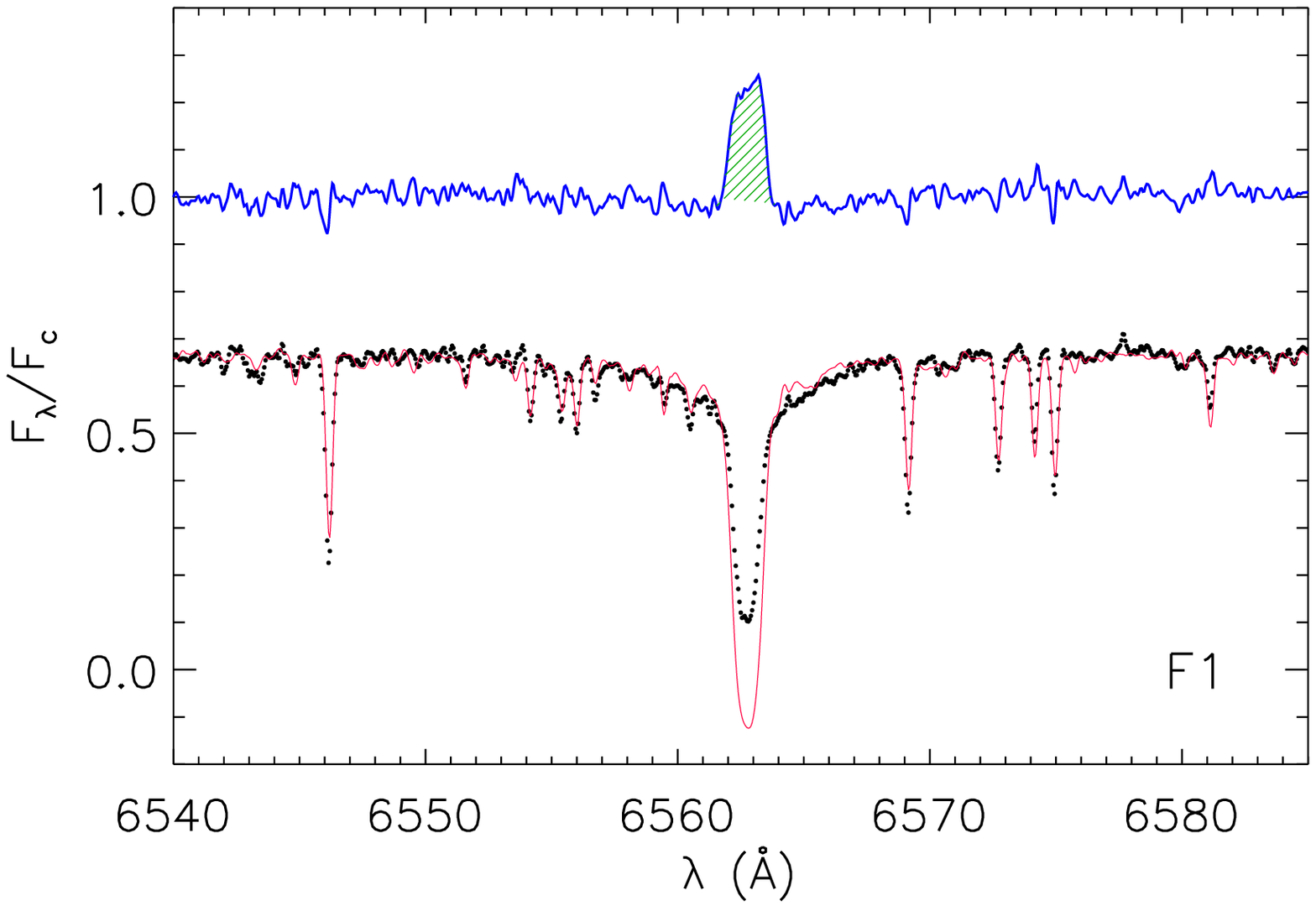}    
\hspace{-.5cm}
\includegraphics[width=5.9cm]{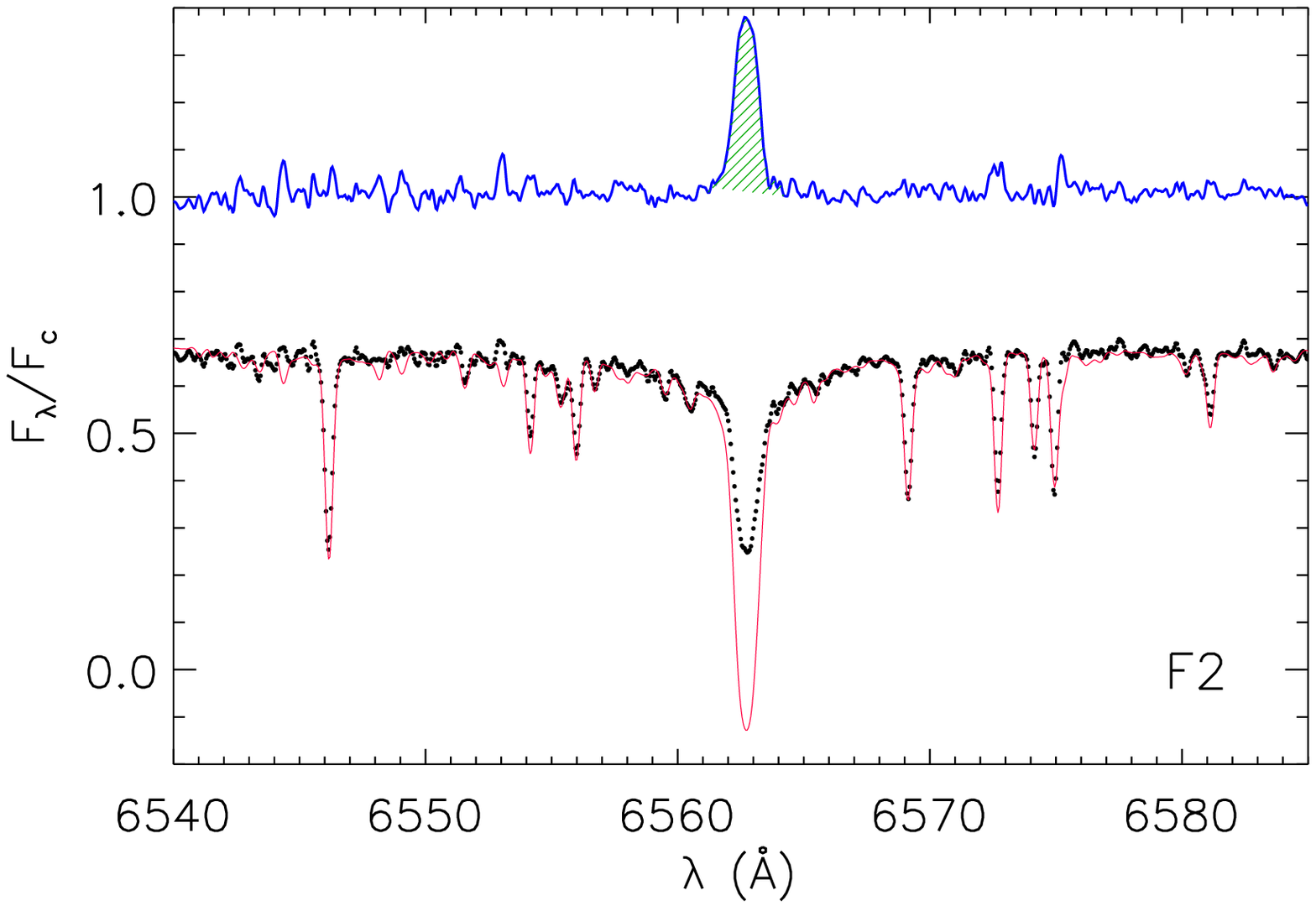}    
\vspace{0cm}
\caption{HARPS-N spectra of the investigated stars in the H$\alpha$ region. 
In each box, the non-active template (red line) is overlaid with the observed spectrum (black dots). 
The chromospheric emission which fills in the H$\alpha$ core is clearly visible 
in the subtracted spectrum (blue line in each panel). The green hatched area represents the excess H$\alpha$
emission that was integrated to obtain $W_{\rm H\alpha}$. The ID of the source is marked in the lower right corner of each box.}
\label{fig:subtraction_halpha}
\end{center}
\end{figure*}

\end{appendix}

\end{document}